\documentclass[fleqn,useAMS,usenatbib]{mnras}

\usepackage[T1]{fontenc}
\usepackage{ae,aecompl}

\usepackage{pdflscape}
\usepackage{graphicx}	
\usepackage{amsmath}	
\usepackage{amssymb}	
\usepackage{color}

\newcommand{\kms}{\hbox{${\rm km~s}^{-1}$}}
\newcommand{\hi}{\ion{H}{i}}

\newcommand{\Msun}{\ensuremath{M_{\sun}}}
\newcommand{\Mstar}{\ensuremath{M_{\star}}}
\newcommand{\logmstar}{\ensuremath{\log \, (M_{\star}/M_{\sun})}}
\newcommand{\logmstarshort}{\ensuremath{\log \, M_{\star}}}
\newcommand{\vrot}{\ensuremath{V_{\mathrm{rot}}}}
\newcommand{\logvrot}{\ensuremath{\log V_{\mathrm{rot}}}}

\newcommand{\deltapa}{\ensuremath{\Delta \mathrm{PA}_{\mathrm{dp}}}}

\newcommand{\fgas}{\ensuremath{f_{\mathrm{gas}}}}
\newcommand{\logfgas}{\ensuremath{\log f_{\mathrm{gas}}}}
\newcommand{\gmr}{\ensuremath{g - r}}

\newcommand{\sfourg}{S\ensuremath{^{4}}G}

\newcommand{\atwomax}{\ensuremath{A_{2,{\rm max}}}}
\newcommand{\afourmax}{\ensuremath{A_{4,{\rm max}}}}

\newcommand{\pbeta}{\ensuremath{P_{\beta=0}}}

\newcommand{\flatprof}{Peak+Shoulders}
\newcommand{\flatp}{P+Sh}
\newcommand{\bpsample}{B/P-Detection Subsample}
\newcommand{\faceon}{Face-on Subsample}

\defcitealias{erwin18}{Paper~I}

\title[Bar Profiles]{The Profiles of Bars in Spiral Galaxies}
\author[P. Erwin et al.]{Peter Erwin$^{1,2}$\thanks{E-mail: erwin@mpe.mpg.de}, 
Victor P. Debattista$^{3}$ and Stuart Robert Anderson$^{3}$ \\
$^{1}$Max-Planck-Insitut f\"{u}r extraterrestrische Physik, Giessenbachstrasse, 85748 Garching, Germany \\
$^{2}$Universit\"{a}ts-Sternwarte M\"{u}nchen, Scheinerstrasse 1, D-81679 M\"{u}nchen, Germany \\
$^{3}$Jeremiah Horrocks Institute, University of Central Lancashire, Preston PR1 2HE, UK}

\date{Accepted XXX. Received YYY; in original form ZZZ}

\pubyear{2023}

\begin{document}
\label{firstpage}
\pagerange{\pageref{firstpage}--\pageref{lastpage}}
\maketitle

\begin{abstract} 

We present an analysis of major-axis surface-brightness profiles of bars
in a volume-limited sample of 182 barred spiral galaxies, using
\textit{Spitzer} 3.6\micron{} images. Unlike most previous studies, we
use the entire bar profile, and we classify profiles into four
categories. These are ``Peak+Shoulders'' (\flatp{}) -- updating the
classic ``flat bar'' profile -- and three subtypes of the classic
``exponential'' profile: (true) Exponential, ``Two-Slope'' (shallow
inner slope + steeper outer slope), and ``Flat-Top'' (constant inner
region, steep outer slope). \flatp{} profiles are preferentially found
in galaxies with high stellar masses, early Hubble types, red colours,
and low gas fractions; the most significant factor is stellar mass, and
previous correlations with Hubble type can be explained by the tendency
of higher-mass galaxies to have earlier Hubble types. The most common
type of non-\flatp{} profile is Exponential, followed by Flat-Top
profiles; all non-\flatp{} profiles appear to have similar distributions
of stellar mass, Hubble type, colour, and gas fraction. We also
morphologically classify the bars of an inclined subsample into those
with and without boxy/peanut-shaped (B/P) bulges; as previously
reported, the presence of a B/P bulge is very strong function of stellar
mass. Essentially all bars with B/P bulges have \flatp{} profiles; we
associate the profile shoulders with the outer, vertically thin part of
the bar. We find a small number of \flatp{} profiles in bars without
clear B/P bulges, which may indicate that \flatp{} formation precedes
the formation of B/P bulges.
\end{abstract}


\section{Introduction}\label{sec:intro} 



The majority of spiral galaxies with stellar masses of
$10^{9}$--$10^{11} \Msun$ have at least one stellar bar
\citep[e.g.,][]{sheth08,dg16a,erwin18}. Bars are found in both blue,
actively star-forming galaxies and in ``red and dead'' systems,
including S0 galaxies, and have been detected in galaxies with redshifts
as high as $\sim 2$ \citep{guo23}. Given this prevalence, we might
wonder if all bars are basically identical, or if they come in different
forms, possibly linked to different host-galaxy characteristics,
formation scenarios, or stages of development. One of the most
fundamental and easily studied characteristics of bars -- setting aside
even more basic measurements like size -- is their \textit{radial
surface-brightness profiles} (a manifestation of their stellar-density
structure), which traditionally come in at least two varieties.

\subsection{The Traditional Picture: Flat and Exponential Bar Profiles}\label{sec:intro-trad}

The classic, pioneering study of bar profiles is that of \citet{ee85},
who studied surface-brightness profiles along the major and minor
axes of bars in 15 barred spiral galaxies. They argued that bar profiles
fell into two classes: ``flat'' and ``exponential''. The distinction was
based on a comparison of the major-axis profile in the bar region to the same
profile \textit{outside} the bar (beyond the radius of the bar,
in what they termed the ``spiral'' region). Flat profiles tended to have
shallow or even constant surface-brightness profiles in the bar region,
with the profile becoming steeper (and exponential) outside the bar.
Exponential profiles, on the other hand, had bar profiles that were
exponential, with slopes \textit{at least as steep as} the profile
outside the bar. They found evidence for a clear difference in Hubble
types: flat profiles were preferentially found in earlier Hubble types,
with exponential profiles in later types. Follow-up studies using
near-IR imaging \citep{elmegreen96,regan97} provided further support for
this dichotomy.

It is important to bear in mind that what \citet{ee85} meant by the term
``flat'' is \textit{not} that the profile must be literally flat (i.e.,
constant surface brightness as a function of radius), though it
certainly does include such cases (as well as extreme cases where the
profile gets brighter towards the end of the bar). They said, ``each
[flat-bar] galaxy has a surface brightness that decreases slowly or not
at all with increasing distance along the bar, and decreases more
rapidly along the spirals.  We refer to these bar profiles as
\textit{flat} (i.e., flat when compared with the spiral profiles; some
of these bar profiles are still `exponential-like,' but their slopes are
\textit{smaller} than the slopes in the spiral regions).'' 

We emphasize this point because some subsequent studies have nonetheless
interpreted the term literally, and then claimed that such profiles are
less common. For example, \citet{seigar98} analyzed near-IR images of 40
barred galaxies and argued that there was no correlation with Hubble
type, in contrast to the findings of \citet{ee85}. Inspection of their
Fig.~14 shows that what they call the ``exponential'' profile of
NGC~5737 is really of the flat type -- though with less broad and
dramatic shoulders than the ``flat'' bars of IC~357 and IC~568 in the
same figure.\footnote{See Section~\ref{sec:intro-bp} and
Figure~\ref{fig:flatbar-BP-demo} for definitions and examples of
``shoulders'' in bar profiles.} \citet{buta06} analyzed the $K$-band bar
profiles of 26 S0--Sa galaxies and introduced an ``intermediate'' class
of profiles, in addition to flat and exponential. Inspection of their
profiles shows this new class has the same basic ``shallow + steep''
shape as in \citet{ee85} -- in fact, two of their
``intermediate-profile'' galaxies (NGC~4596 and NGC~4608) were
classified as flat by Elmegreen \& Elmegreen. 

More recent studies have focused on identifying flat versus exponential
profile via 2D fits to galaxy images, using disc + bar + bulge models
where the bar is a 2D elliptical structure with a S\'ersic radial
surface-brightness profile. In this approach, the ``flatness'' of the
bar profile is represented by the S\'ersic index $n$, with lower values
corresponding to flatter profiles.\footnote{We remind the reader that
the S\'ersic profile is an exponential when $n = 1$ and a Gaussian when
$n = 0.5$.}  \citet{kim15} modeled \textit{Spitzer} IRAC1 (3.6\micron)
images of 144 low- and moderate-inclination barred galaxies from the
\textit{Spitzer} Survey of Stellar Structure in Galaxies
\citep[][\sfourg]{sheth10}; they considered bar S\'ersic indices $n <
0.4$ (i.e., sub-Gaussian) as representative of flat bars. Their main
finding was that low-$n$ bars were preferentially found in high-mass
galaxies, with a transition mass of $\logmstar \sim 10.2$. They also
argued that an even better separation could be had using the $B/T$ value
from their fits, with $B/T > 0.2$ having almost exclusively flat bars
and ``bulgeless'' ($B/T = 0$) galaxies having almost exclusively
exponential-like bars. \citet{kruk18} performed 2D fits for 3461 barred
galaxies using SDSS images and found a similar mass-based pattern:
barred galaxies with $\logmstar \geq 10.25$ had a mean $n_{\rm bar} =
0.43$, while lower-mass galaxies had a mean $n_{\rm bar} = 0.81$ (nearly
exponential). They, too, noted an association between prominent bulges
and flatter bars, though this was based on the presence or absence of a
bulge component in their best-fitting 2D models: ``disc-dominated''
galaxies (where only a disc and bar components were needed in the model)
had $n_{\rm bar} = 0.92$, while ``obvious bulge'' galaxies (where a
bulge component was needed in addition to the disc and bar) had $n_{\rm
bar} = 0.40$. However, they did not find a correlation between $B/T$ and
$n_{\rm bar}$ within the ``obvious bulge'' subsample itself.

Relatively little theoretical attention has been paid to the question of
why a dichotomy in bar profiles might exist. \citet{noguchi96} argued
from $N$-body simulations that ``spontaneous'' bars (formed via disc
instabilities) had steep exponential profiles, while bars triggered by
tidal interactions (in discs which were too hot to form bars
spontaneously) had much flatter profiles, complete with ``shoulders'' at
the ends of the bar. However, \citet{athanassoula02a} found that both
types of bar profiles could be found in spontaneously formed bars.
\citet{combes93} argued from their $N$-body simulations that flat-type
profiles were associated with ``early-type'' mass distributions
(centrally concentrated, large bulge/total mass ratios, often with an
inner Lindblad resonance) and bars that extended to corotation, whereas
bars in ``late-type'' mass distributions retained the exponential
profile of the disc. Very recently, \citet{anderson22} studied bar
profiles in a number of $N$-body simulations, finding evidence that bar
profiles transitioned from exponential-like to flat-type over time (and
sometimes transitioned \textit{back}) as part of the bars' secular
evolution. (\citealt{beraldo-e-silva23} investigates the role of bar
resonances in the evolution of these features.)

We believe there are two reasons why now is a good time to revisit the
question of bar profiles. The first is that past observational studies
almost always involved relatively small, heterogeneous samples, making
it difficult to draw firm conclusions about the prevalence of different
profile types and their possible relation to other bar and galaxy
parameters (partial exceptions to this trend are \citealt{kim15} and
\citealt{kruk18}, although they focused on fitting with simple 2D
functions rather than detailed nonparametric analysis of profiles).
There now exist sufficient high-$S/N$ near-IR images with decent spatial
resolution -- in particular, as part of \sfourg{} -- for larger,
unbiased samples of barred galaxies to be constructed.

Second, recent theoretical and observational studies have given us a
much clearer understanding of what flat bar profiles are really like,
as we discuss in the following subsection.

\subsection{New Insights: The Role of Boxy/Peanut-Shaped Bulges in Bar Profiles}\label{sec:intro-bp}

A significant body of work
\citep[e.g.,][]{combes90,raha91,kuijken95,bureau99a,
athanassoula02a,athanassoula05b,mendez-abreu08,erwin-debattista13,wegg15,herrera-endoqui17,blana-diaz17} 
has demonstrated that many bars consist of two distinct,
three-dimensional stellar components: an outer region which is
vertically thin and an inner region which is vertically thick, with a
``boxy'' or ``peanut-shaped'' appearance when seen from the side (i.e.,
in edge-on galaxies with the bar oriented close to perpendicular to the
line of sight). This vertically thick inner part of the bar is thus
usually referred to as the box/peanut or B/P bulge. When seen at
intermediate inclinations, the B/P projects to form a broad, sometimes
boxy, shape in the isophotes (the ``box'' in the terminology of
\citealt{erwin-debattista13}), while the outer, vertically thin part of
the bar appears as thin, offset ``spurs'' extending to larger radii. In
systems close to face-on, the B/P bulge can still be identified as the
rounder, inner part of the bar (a ``barlens'', in the terminology of
Laurikainen and collaborators). The studies of
\citet{erwin-debattista17}, \citet{li17}, \citet{kruk19}, and
\citet{marchuk22} found that such two-component bars are preferentially
found in higher-mass galaxies, and thus in early-type spirals and S0s.
This means that bars with B/P bulges appear to be found in the same
types of galaxies that have flat bar profiles.

This connection is important because modern investigations of B/P bulges
in both real galaxies and simulations have shown that the B/P-bulge
region has a distinct, \textit{steeper} surface-brightness profile
compared to that of the outer part of the bar
\citep{laurikainen14,athanassoula15}. This suggests that in bars with
flat surface-brightness profiles, the inner part of the bar -- interior to
the classic flat part of the profile -- should
actually have a \textit{steep} profile. Inspection of the profiles of
classic flat-bar galaxies \citep[e.g.,][]{ee85,elmegreen96, regan97}
shows that this is probably the case. (In past studies, the inner parts
of the bar profile were ignored, either because they were in a saturated
part of the image or possibly because they were considered to belong to
the ``bulge'' and were thus not considered part of the bar.)

\begin{figure*}
\begin{center}
\hspace*{-3.5mm}\includegraphics[scale=0.9]{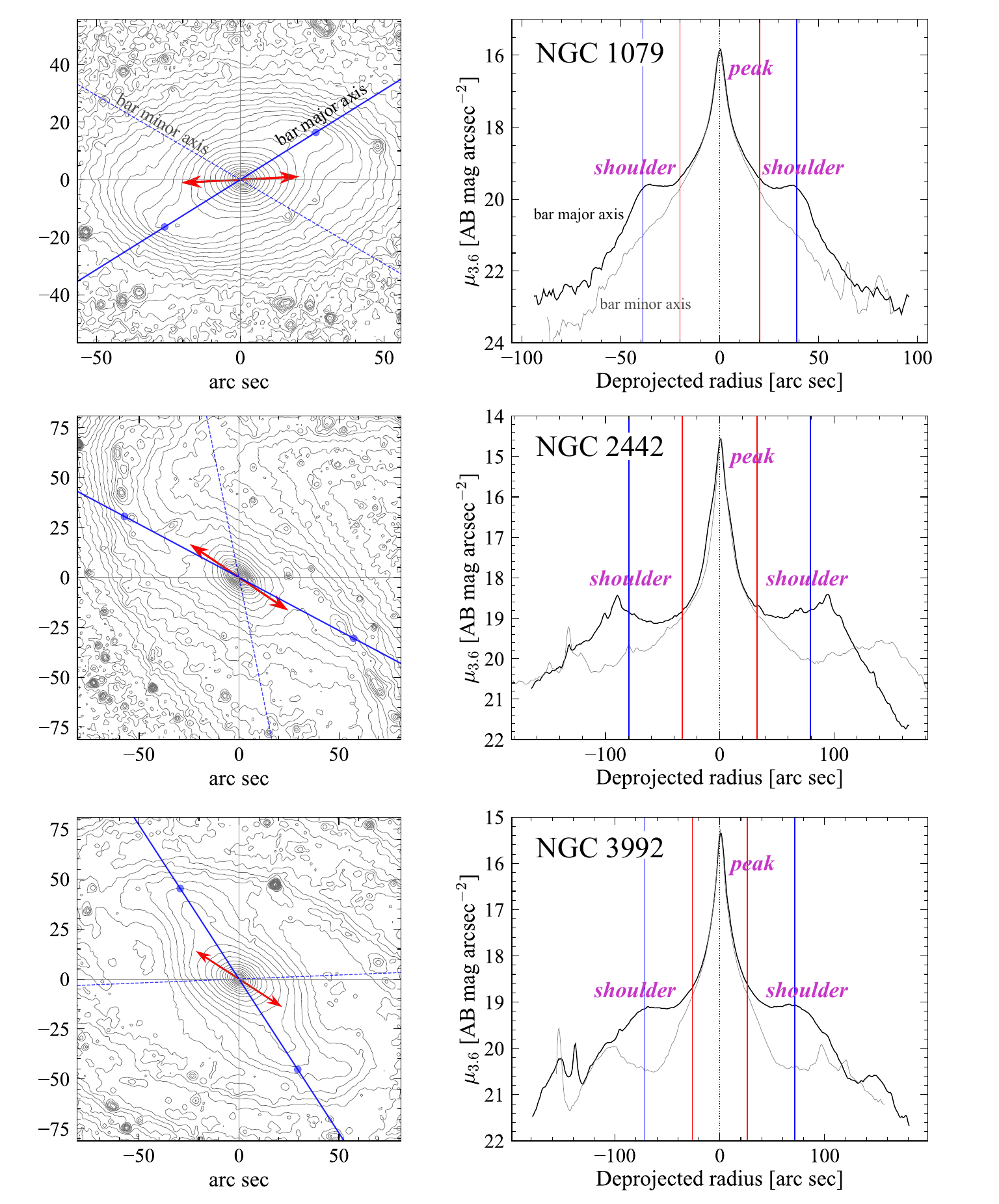}
\end{center}

\caption{Plots of bar isophotes and profiles for three classic ``flat
bar'' galaxies, along with examples of the ``peak'' and ``shoulders''
sub-components: NGC~1079 \citep{buta06}, NGC~2442 \citep{ee85}, and
NGC~3992 \citep{regan97}. \textbf{Left:} Logarithmically spaced
isophote contours from \textit{Spitzer} 3.6\micron{} images, showing the bar
region. Thick diagonal blue lines indicate bar position angle, with
blue dots marking approximate bar radius, while thinner, dashed blue lines
indicate the bar minor axis (as projected onto the sky); red arrows
indicate position angle and approximate extent of B/P bulges in each
galaxy (measurements from
\citealt{erwin-debattista13,erwin-debattista17}). \textbf{Right:}
Profiles along bar major axis (thick black lines) and bar minor axis
(thin grey lines), plotted against deprojected radius. Vertical red and
blue lines mark B/P-bulge and bar radius, respectively; magenta labels
indicate the approximate peak and shoulders sub-components of each
profile. Note that the B/P-bulge radius delimits the steep inner peak of
the profile, while the flatter shoulders extend to (and slighty beyond)
the full bar radius.
\label{fig:flatbar-BP-demo}}
\end{figure*}

Figure~\ref{fig:flatbar-BP-demo} shows three examples of galaxies
previously classified as having flat bar profiles which have
\textit{also} been identified as having the morphological
characteristics of B/P bulges. In all three cases, the region of the
B/P bulge (spanned by the red arrows in the left panels and bounded by
vertical red lines in the right panels) is associated with a much
steeper surface-brightness profile, in contrast to the shallower
profile outside, associated with the thinner and misaligned bar spurs.

Given our modern understanding of bars, then, we can argue that the
classic flat-bar profile is better understood as a \textit{two-part}
profile. The inner region is steep and has usually been considered to be
part of the (classical, spheroidal) bulge (as has been done with recent
2D fitting of barred-galaxy images by, e.g., \citealt{kim15} and
\citealt{kruk18}), instead of actually being the inner part of the bar.
The outer part is the ``flat'' (or shallow) region, which is also
marked, at the very end of the bar, by a break and a steep outer falloff
that leads to the disc outside the bar. We suggest the term
\textbf{shoulder} for the full outer part of the profile: the interior shallower
profile plus the break and falloff outside.\footnote{This
terminology is not original to us; earlier uses include, e.g.,
\citet{noguchi96} for $N$-body bars and \citet{gadotti07} for real bars.
This is also the same as what \citet{athanassoula06} termed a ``hump''
in the profiles of M31's bar and of some $N$-body bars.} We call
the entire profile -- the combination of the shoulder with the steep
inner part of the profile (with a central maximum) --
\textbf{\flatprof}, or \textbf{\flatp} for short.

\subsection{Moving Forward}

If the traditional flat bar profile is better understood as a two-part
structure, could something similar be the case for exponential bar
profiles? After all, these were also defined in \citet{ee85} on the
basis of only the outer parts of the bar (due to centrally saturated
images), so it would be useful to know if they are truly, as the name implies,
a single exponential profile that extends all the way in to the centre.

Our approach in this paper is to take a comprehensive look at local bar
profiles using modern data and a relatively unbiased, complete sample of
galaxies. Specifically, we use near-IR imaging data from
\textit{Spitzer}, which minimizes possible confusion from dust and star
formation,\footnote{We checked that hot dust and PAH from star formation
does not significantly affect our profile classifications; see
Section~\ref{sec:data-sources}.} and we focus on the \textit{entire}
bar-major-axis profile, from the centre of the galaxy through the end of
the bar. We do this for \textit{all} bars in a volume- and mass-limited
sample of spiral galaxies (not including lenticular galaxies). (In the
future, we also plan to compare our observational findings with the
results of analyzing bar profiles in barred galaxy simulations, as in
\citealt{anderson22}.)

Among the questions we hope to address are: Are all \flatp{} profiles
due to the presence of a B/P bulge? What do the central regions of exponential
profiles look like? Can we identify any additional classes of bar profiles?
Is there a more fundamental (physical?) driver of bar-profile type than
Hubble type?

\section{Sample Definitions and Data Sources}\label{sec:samples} 

\subsection{Sample Definitions} 

We started, following the general philosophy of \citet{erwin18} and
\citet{erwin19}, with subsamples of the \citet{dg16a} subset of \sfourg:
all disc galaxies in \sfourg{} with inclinations $i \leq 65\degr$. We
removed galaxies that lack distances and
stellar masses in the compilation of \citet{munoz-mateos15}, galaxies
without reliable distances (i.e., radial velocities of $< 500$ \kms{} and no
non-redshift-based distance estimates), and galaxies with optical
diameters below the formal \sfourg{} limit of $D_{25} = 1\arcmin$. We
also restricted the sample to spiral galaxies, due to the strong bias
against S0 galaxies in \sfourg. This yielded the ``Parent Spiral Sample'',
with 1220 galaxies.\footnote{This is the same as the ``Parent Spiral
Sample''' of \citet{erwin18}, but slightly different from the ``Parent Spiral
Sample'' of \citet{erwin19}.} (In the future, we plan to expand this 
analysis to include S0 galaxies.)

\begin{figure*}
\begin{center}
\hspace*{-3.5mm}\includegraphics[scale=0.9]{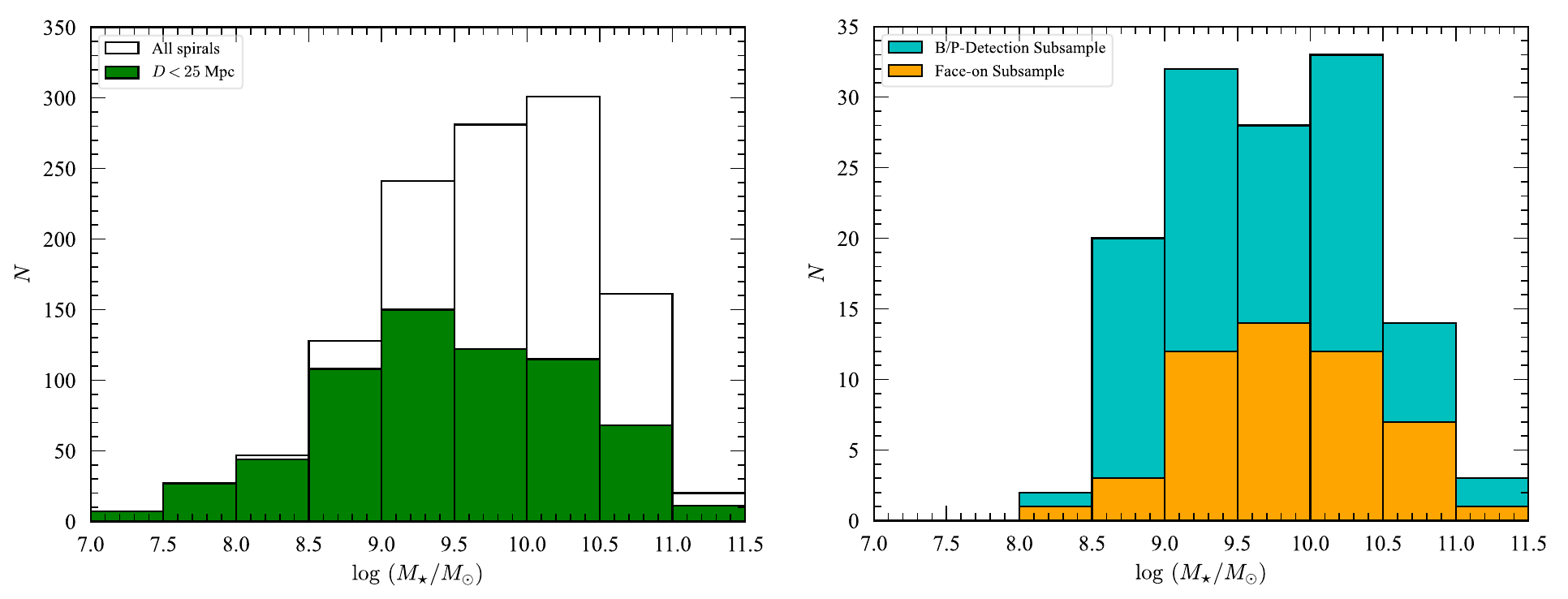}
\end{center}

\caption{Left: Stellar-mass distributions for spiral galaxies in the Parent
Spiral Sample (hollow) and the $D < 25$ Mpc Main Spiral Sample (green).
Right: Same, but now for the two barred-galaxy subsamples (\bpsample{} [cyan]
and \faceon{} [orange]) analyzed in this paper.\label{fig:sample-mstar}}

\end{figure*}

We then applied a distance limit of 25 Mpc. This ensured
good completeness down to a stellar mass of $\logmstar \sim 8.5$, as
well as reasonable spatial resolution (median point-spread-function FWHM
$\sim 165$ pc), such that most bars can be successfully detected; see
the discussion in \citet{erwin18}. This reduced the sample to 659
spirals (``Main Spiral Sample''), of which 370 have bars according to
the catalog of \citet{herrera-endoqui15}. The left panel of
Figure~\ref{fig:sample-mstar} shows the stellar-mass distributions of
the Parent and Main Spiral Samples.

We created two subsamples to analyze. The first -- the
\textbf{\bpsample} -- was defined by selecting those galaxies with
orientations best suited to allow us to determine if the bars did or did not
have B/P bulges; the rationale was to allow us to
test possible associations between B/P bulges (or the lack thereof) and
bar profiles. Following the precepts of \citet{erwin-debattista17}, we
selected galaxies with inclinations $> 40\degr$ and relative bar
position angles $\deltapa$ (deprojected angle between bar and disc major
axis) $< 60\degr$. This yielded an initial set of 195 galaxies. We
subsequently discarded 52 galaxies which proved, on closer inspection,
to have dubious or nonexistent bars (typically in low-mass, late-type
galaxies), along with five galaxies with erroneous inclinations (e.g.,
actually edge-on or face-on) and seven galaxies where bright stars close
to or superimposed on the galaxies would interfere with the extraction
of clean bar-major-axis profiles. (We also added one galaxy
originally placed in the \faceon; see next paragraph.) This left us
with a total of 132 galaxies in the \bpsample.

The second subsample -- the \textbf{\faceon} -- was meant to focus on
\textit{low-inclination} galaxies. We intended this as a comparison to
the \bpsample, to see if differences in inclination might have any
effects on our profile classifications. We selected galaxies from the
Main Spiral Sample with inclinations $\leq 30\degr$, with no
restrictions on bar orientation. After discarding six galaxies with
dubious or nonexistent bars and two galaxies actually too highly
inclined,\footnote{One galaxy had an inclination of $\sim32\degr$;
the other had both inclination and bar orientation appropriate for the
\bpsample, so we added it to the latter.} we were left with 50 galaxies
in the \faceon. Taken together, our final combined subsamples comprise
182 galaxies.

The right panel of Figure~\ref{fig:sample-mstar} shows the stellar-mass
distributions of both subsamples, which show no signs of being
significantly different (Kolmogorov-Smirnov [K-S] test $P = 0.19$). The strong
decline for $\logmstar < 9$ is due in part to the low frequency of
barred galaxies among low-mass galaxies \citep{erwin18}.

\subsection{Data Sources}\label{sec:data-sources} 

Our primary data source was the \textit{Spitzer} IRAC1 (3.6\micron)
images of the \sfourg{} sample. We used these for two purposes:
\begin{enumerate}
\item Extraction of the bar-major-axis and bar-minor-axis profiles;
\item Identification of B/P-bulge morphologies.
\end{enumerate}

Images were background-subtracted by taking the mean of several dozen median
measurements in $20 \times 20$-pixel boxes in relatively blank regions of 
the image well outside the galaxy. We note that accurate background
subtraction is not essential for our purpose, since we are interested in
the isophote shapes and surface-brightness profiles in the bright central
regions of the galaxies.

Bar sizes and bar orientations were taken from \citet{herrera-endoqui15};
in 26 cases, we revised the bar position angles to better match the
images. Bar strengths, when available, were taken from
\citet{dg16a}.\footnote{Of the 182 galaxies in our combined subsamples,
147 have deprojected ellipticity values, 142 have \atwomax{} values, and
105 have \afourmax{} values.} Galaxy centre (pixel) coordinates, along
with disc orientations (position angle of the major axis and
inclination), were taken from \citet{salo15}; we determined revised
centres for 35 galaxies. For one galaxy (IC~796), we adopted a different
disc position angle (144\degr, versus the 137\degr{} of Salo et al.)
Stellar masses were taken from \citet{munoz-mateos15}, while neutral gas
masses were based on the m21c value in HyperLEDA, as described in
\citet{erwin18}.\footnote{For NGC~4314,
which has no m21c value in HyperLEDA, we use the \hi{} flux in
\citet{springob05}.} Gas rotation velocities \vrot{} were taken from
HyperLEDA, using their $W_{\rm gas}$ parameter corrected for
inclination. Finally, global \gmr{} galaxy colours were based on the 2020
version of the Siena Galaxy Atlas
(SGA-2020)\footnote{\url{https://www.legacysurvey.org/sga/sga2020/}} --
specifically, the total curve-of-growth $g$ and $r$ magnitudes
(``MTOT''). These are not available for 20 galaxies in the combined
subsamples.

Detailed analysis of \sfourg{} images has suggested that when star
formation is present, as much as $\sim 10$--30\% of the emission in the
IRAC1 filter can be due to hot dust and PAH emission rather than old
stars \citep{meidt12,querejeta15}. To check how much this might
contaminate our profiles, we downloaded ``stellar'' images from IRSA 
\citep[as produced by ``Pipeline 5'' in][]{querejeta15} for
a subset of our samples and generated bar-major-axis profiles from them
to compare with the profiles from the original IRAC1 images.
(Only $\sim 55$\% of our galaxies have stellar images from Pipeline 5.)
Although there are localized differences, the overall forms of the
profiles do not change meaningfully between the IRAC1 and stellar images,
and thus our classifications do not change.

\section{Classification of Bar Profiles}\label{sec:profile-classif} 

\subsection{The Four Types of Bar Profiles} 

Our starting point for bar-profile classifications was the original
\citet{ee85} ``flat'' and ``exponential'' classes. As discussed
in Section~\ref{sec:intro}, we recognize that ``flat'' profiles are only
shallow or flat in their outer parts; in most if not all cases, these
bars have steep inner profiles, which appear to be at least partly due
to the inner, B/P structure of the bar. Because of this, we departed from
\citet{ee85} and most subsequent work by considering the \textit{entire}
profile, from the centre of the galaxy ($r = 0$) out to the end of the
bar, even though this will inevitably include some contribution in the
central part of the profile from non-bar components such as classical
bulges, nuclear discs, nuclear star clusters, etc. Consequently, we
began by replacing the term ``flat'' with the term \textbf{\flatprof{}
(\flatp)}; this reflects both the full outer part of the bar profile
(the shoulder) and the steep inner component (the peak).

\begin{figure*}
\begin{center}
\hspace*{-3.5mm}\includegraphics[scale=0.9]{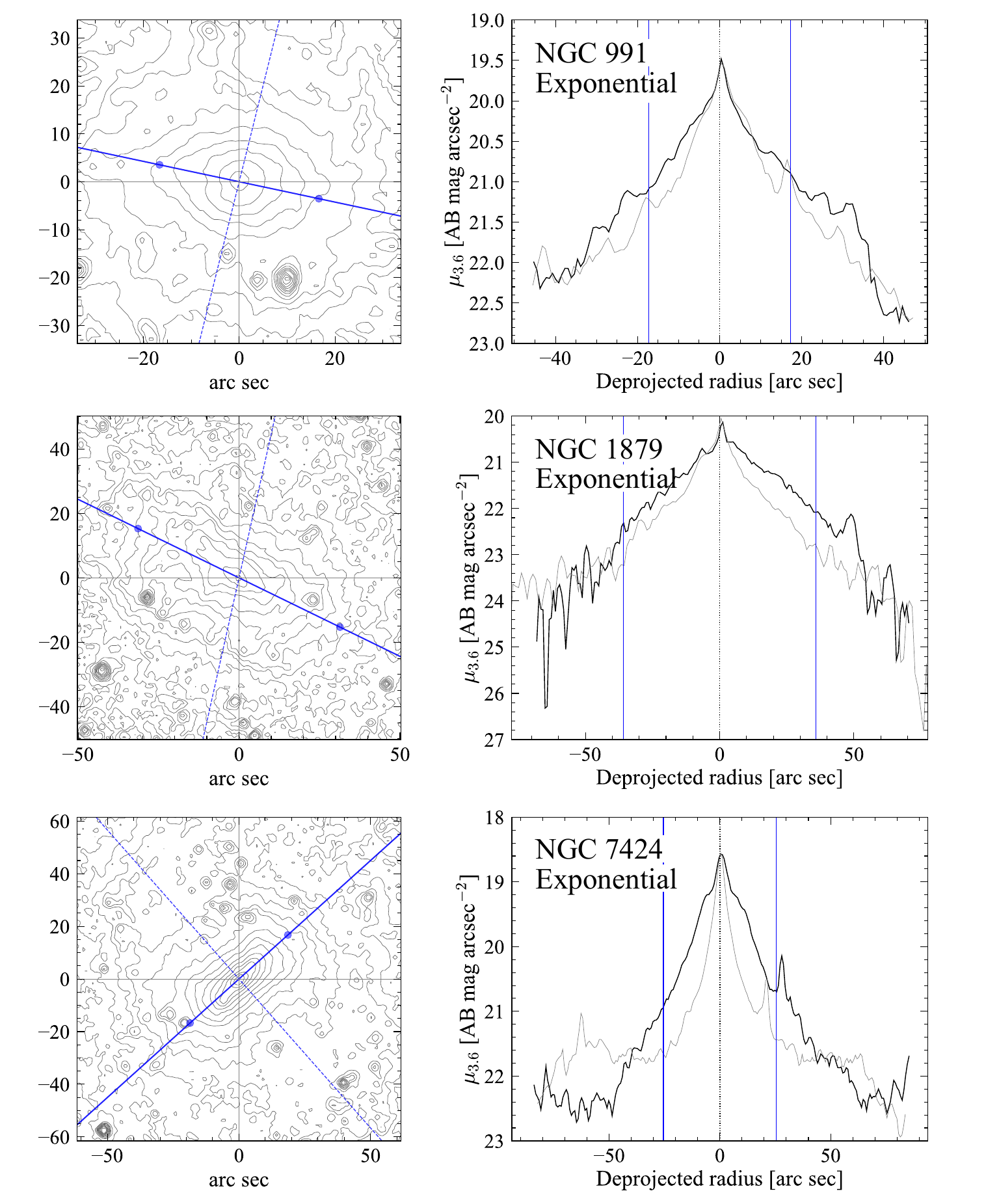}
\end{center}

\caption{As for Figure~\ref{fig:flatbar-BP-demo}, but showing examples
of galaxies with Exponential (Exp) bar profiles. Left: \textit{Spitzer}
3.6\micron{} isophotes. Thick (solid) and thin (dashed) blue lines
indicate bar major-axis and minor-axis position angles, respectively (as
projected onto the sky); blue circles mark the approximate radius of the
bar. Right: Profiles along bar major axis [solid lines] and minor axis
[thin grey lines]. Vertical blue lines indicate bar radius
\label{fig:exp-demo}}

\end{figure*}

\begin{figure*}
\begin{center}
\hspace*{-3.5mm}\includegraphics[scale=0.9]{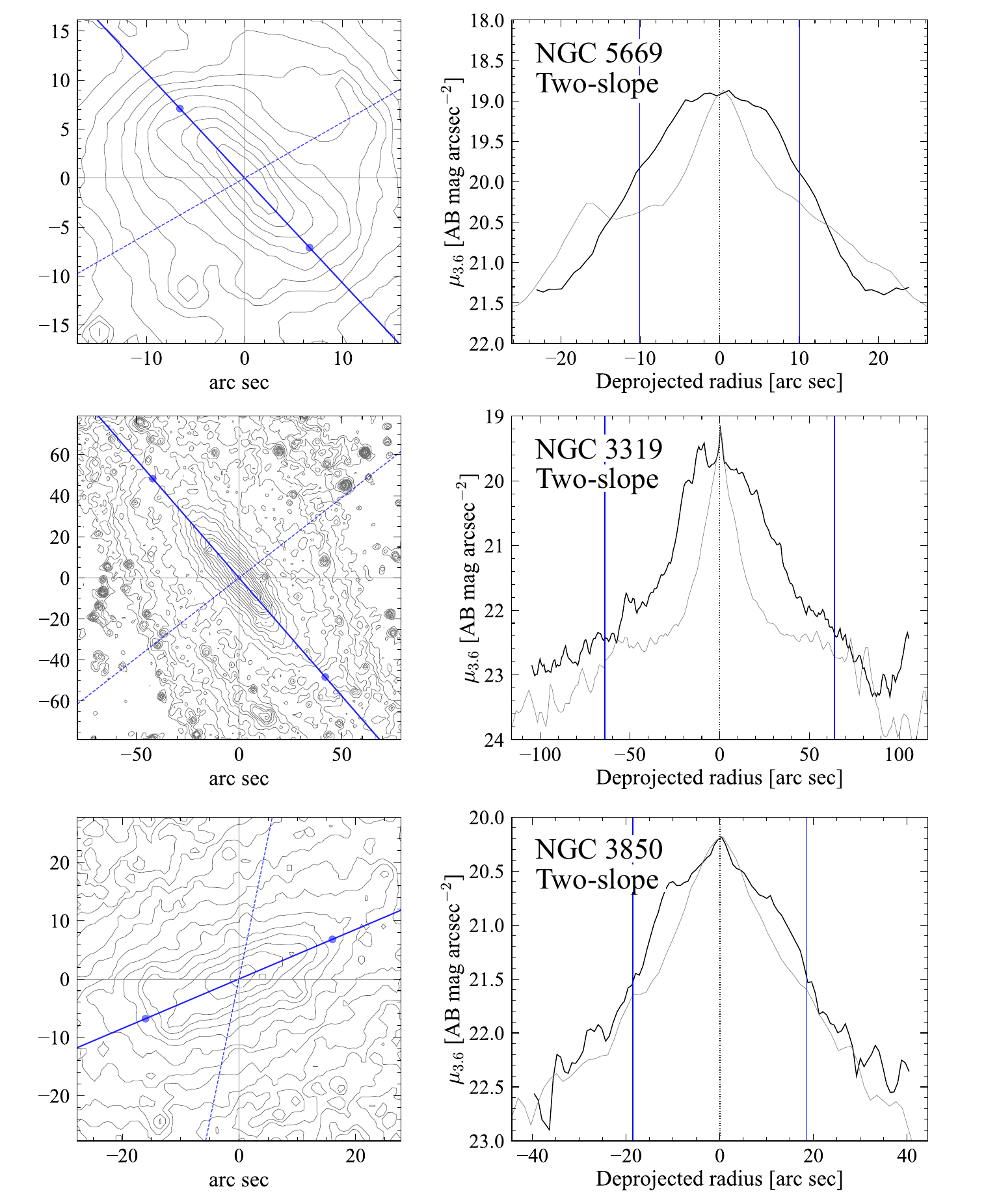}
\end{center}

\caption{As for Figure~\ref{fig:exp-demo} but now showing examples of
galaxies with Two-Slope (2S) bar profiles. Note that NGC~3319 has a small
nuclear peak, consistent with a nuclear star cluster \citep[e.g.,][]{georgiev14};
NGC~3850 may have a weaker example. \label{fig:two-slope-demo}}
\end{figure*}

\begin{figure*}
\begin{center}
\hspace*{-3.5mm}\includegraphics[scale=0.9]{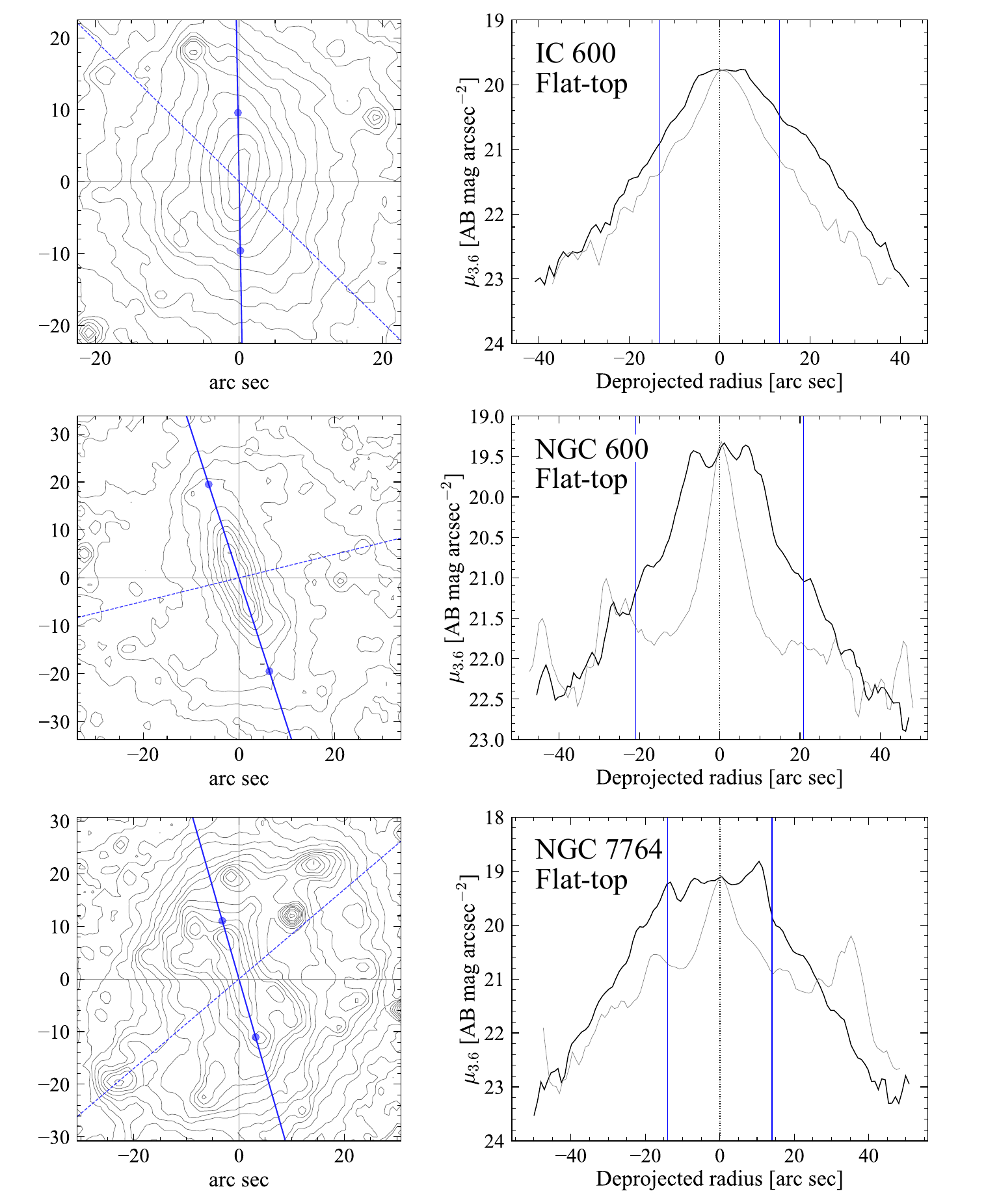}
\end{center}

\caption{As for Figure~\ref{fig:exp-demo} but now showing examples of
galaxies with Flat-Top (FT) bar profiles. \label{fig:flattop-demo}}

\end{figure*}

Given this approach, we also considered the entire profile ($r \geq 0$)
of ``exponential'' bars. Exploratory analysis of bar-major-axis profiles
from our sample suggested that when this is done, there is actually more
than one type of ``exponential'' profile. That is, among the profiles
which were \textit{not} \flatp, we found many with genuinely
single-exponential profiles, but also a number with more complicated
forms. We eventually settled on a total of \textit{three} classes of
non--\flatp{} bar profiles.
Figures~\ref{fig:exp-demo}--\ref{fig:flattop-demo} show examples of
these three types of bar profiles. Note that some of these profiles do
show compact central peaks, plausibly consistent with nuclear star
clusters; we note the existence of such features, but otherwise
ignore them in our analysis.

In summary, we found evidence for \textit{four} general classes of bar profile:

\begin{enumerate}

\item \textbf{\flatprof{} (\flatp)} -- This is our updated version of the classic ``flat''
profile of \citet{ee85}; it includes the ``intermediate'' class of \citet{buta06}.
See Figure~\ref{fig:flatbar-BP-demo} for examples.

\item \textbf{Exponential (Exp)} -- These are profiles which approximate a single
exponential over their whole range, with the possible addition of a small
central excess compatible with a very compact bulge or a nuclear star cluster.
See Figure~\ref{fig:exp-demo} for examples.

\item \textbf{Two-Slope (2S)} -- These are profiles where the inner part is a shallow
exponential, breaking beyond a certain radius (either inside the bar or sometimes
at the bar end)
to a steeper exponential. See Figure~\ref{fig:two-slope-demo} for examples.

\item \textbf{Flat-Top (FT)} -- This can be thought of as an extreme version of the
previous profile: the inner part has essentially constant surface brightness,
breaking to an exponential profile beyond a certain radius. This is distinguished
from the \flatp{} profile by the fact that it has no central ``peak''.\footnote{That is,
these are profiles with shoulders, but no ``heads'' -- headless or decapitated profiles.
(We considered referring to these as ``Green Knight'' profiles, but decided against
such an obscure literary reference.)}
See Figure~\ref{fig:flattop-demo} for examples.

\end{enumerate}

\subsection{How We Did the Classifications} 

Having defined a preliminary classification scheme (see previous
section), we then performed a general visual classification of all 182
bar-major-axis profiles in the two subsamples. To minimize the
possibility of biased classifications, we used a blinded approach where
profile plots were automatically generated and assigned numbers from a
randomized ordering. These profiles were then classified by two of the
authors (PE and VPD). By randomizing the profiles and obscuring their
origins, we hoped to avoid possible biases that might result from
knowing the individual-galaxy origins of profiles (e.g., recognizing a
galaxy name as that of a classic, well-studied ``flat'' or
``exponential'' profile from the literature), knowledge of whether a
profile was from a galaxy with an identified B/P bulge, knowledge of a
galaxy's likely stellar mass,\footnote{In practice, there is the
possibility that one could guess that a very noisy profile came from a
low-mass, low-surface-brightness galaxy.} etc. See
Appendix~\ref{app:plots-for-classif} for examples of the actual plots
used for the blind classification process.

The agreement between the two classifiers was reasonably good, with
disagreements on 24 of the 182 galaxies. The majority of these
disagreements involved exponential sub-types (e.g., Exponential versus
Flat-Top); only eight involved one classifier choosing \flatp{} and the
other choosing one of the exponential sub-types. In general, we count
profile types only for galaxies where both classifiers agreed, though
for overall fractions of profile types we also report results based on
giving each vote for classification X a value of 1 and then dividing the
totals by two. (Thus, a disagreement counts as 1/2 a vote for each of the
two profile types.)

\section{Results for Profile Classifications}\label{sec:profile-results} 

\subsection{General Results}\label{sec:gen-results} 

The two most common profile types we find in our galaxies are \flatp{}
($31^{+4}_{-3}$\% of the combined subsamples) and Exponential ($41 \pm
4$\%); Flat-Top profiles are the next most common ($12^{+3}_{-2}$\%),
with only a small number of bars showing Two-Slope profiles
($3.3^{+1.6}_{-1.1}$\%), along with $12^{+3}_{-2}$\% split
classifications.\footnote{If we count split classifications as 1/2 for each,
then the frequencies are $33^{+4}_{-3}$\% for \flatp, $45 \pm 4$\% for
Exponential, $17 \pm 3$\% for Flat-Top, and $4.7^{+1.8}_{-1.4}$\% for
Two-Slope.} What is potentially more interesting than these raw
percentages is whether the profiles type depends on particular galaxy
characteristics.

\begin{figure*}
\begin{center}
\hspace*{-3.5mm}\includegraphics[scale=0.8]{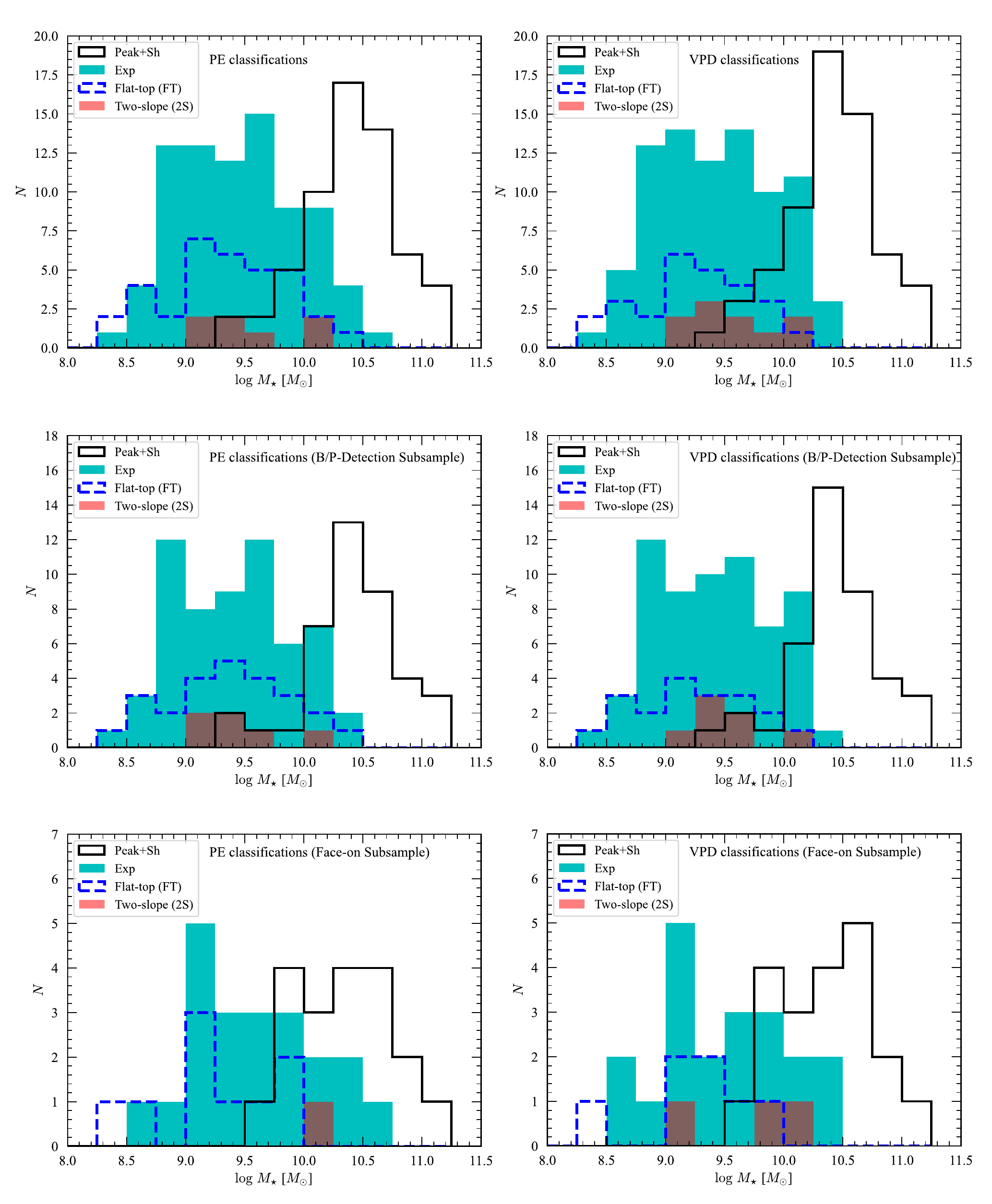}
\end{center}

\caption{Distribution of bar-profile classifications as a function of galaxy
stellar mass, as determined by the two classifiers. Top row: classifications
for all spiral galaxies. Middle row: Classifications for galaxies with moderate
inclinations and low \deltapa{}. Bottom row: Classifications for galaxies
with near-face-on orientations.
\label{fig:hist-profiles-vs-mstar-spirals}}

\end{figure*}

We show the distribution of bar-profiles types as a function of stellar
mass in Figure~\ref{fig:hist-profiles-vs-mstar-spirals}. Results from
the two different classifiers appear in the left and right columns,
respectively; these show very good agreement. The top row has
classifications for all galaxies in both subsamples, while the middle
and bottom rows show the classifications for the \bpsample{} and
\faceon{}, respectively. We see no significant differences for the two
subsamples (K-S $P = 0.50$--0.88), which suggests that inclination does
not meaningfully affect the appearance of the bar profiles.

The clearest trend is the mass segregation between the \flatp{} bars and
the others. \flatp{} bars are preferentially found in high-mass
galaxies, especially $\logmstar \ga 10$. The other profile types, which
are dominated by Exponential bars, are preferentially found in
lower-mass galaxies, and seem to share the same basic
distribution in stellar mass. (A K-S test comparing the stellar mass
distributions of Exponential and Flat-Top profiles yields $P = 0.77$ for
the PE classifications and 0.39 for the VPD classifications, so there is
no evidence for a difference in their parent samples.)

\begin{figure}
\begin{center}
\hspace*{-3.5mm}\includegraphics[scale=0.8]{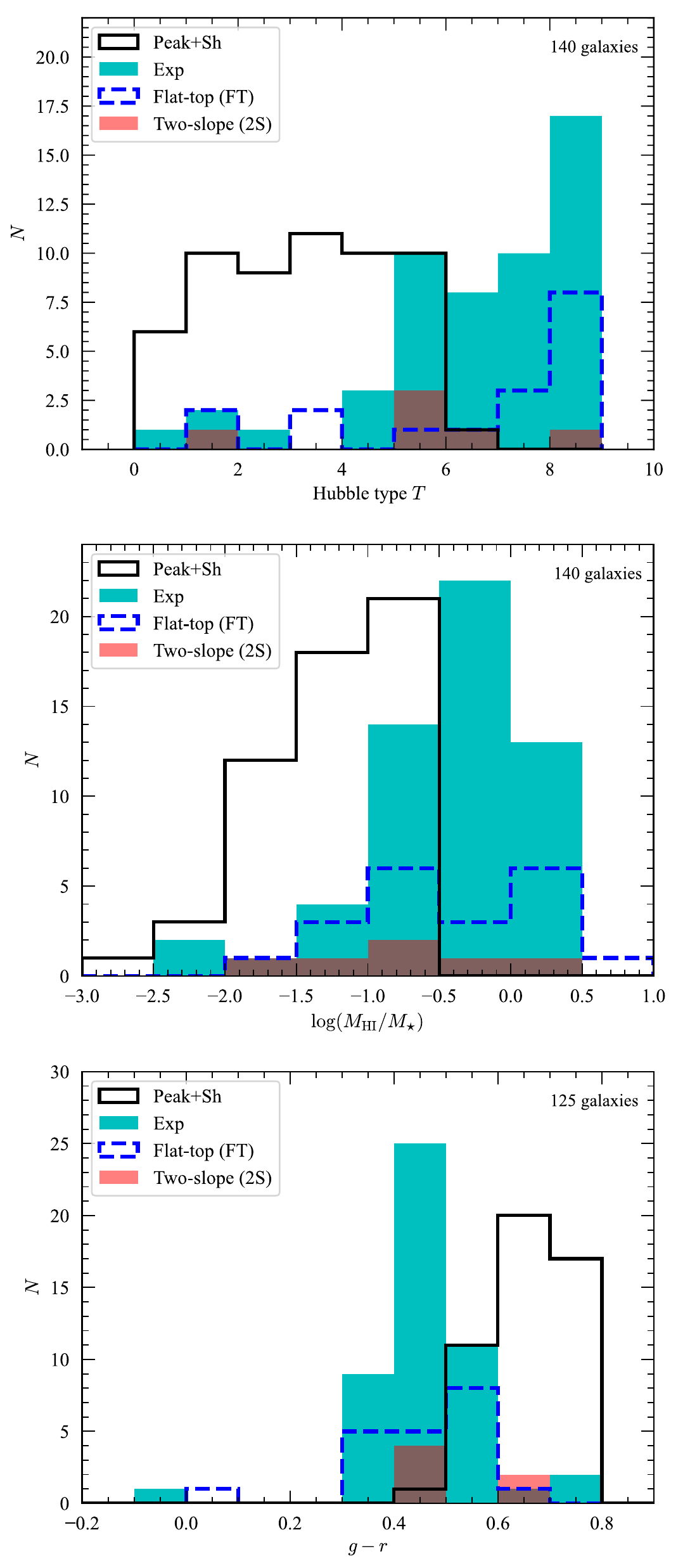}
\end{center}

\caption{Distribution of bar-profile classifications as a function of
galaxy Hubble type $T$ (top), \hi{} mass fraction (middle), and \gmr{}
colour (bottom), as determined by the two classifiers for all spiral
galaxies. In contrast to
Figure~\ref{fig:hist-profiles-vs-mstar-spirals}, here we show
distributions for galaxies where both classifiers agree on the profile
type, combining both subsamples. The number of galaxies are indicated in the
upper-right corners of each plot (some galaxies do not have \gmr{}
colours).
\label{fig:hist-profiles-vs-htype-fgas-gmr-spirals} }

\end{figure}

Figure~\ref{fig:hist-profiles-vs-htype-fgas-gmr-spirals} shows the
distributions of Hubble types (top),  neutral gas mass fraction
(middle), and \gmr{} colour for each profile type, with classifications
from both PE and VPD merged. As was originally pointed out by
\citet{ee85}, \flatp{} profiles (their ``flat'' profiles) are typically
found in earlier Hubble types, while the Exponential and other profiles
become common only for Sc and later galaxies. We can also see that
gas-poor galaxies are more likely to have \flatp{} profiles, while
gas-rich galaxies tend to have a mix of the non-\flatp{} profiles. A
similar trend holds for colour, with red galaxies more likely to have
\flatp{} profiles. This is consistent with the stellar-mass and
Hubble-type trends, in that high-mass galaxies (and earlier Hubble
types) tend to be redder and more gas-poor than low-mass galaxies (and
later Hubble types). We examine the question of whether there is an
\textit{independent} relation between profile type and any of these
three parameters in Section~\ref{sec:multiple-params}.

\begin{figure*}
\begin{center}
\hspace*{-3.5mm}\includegraphics[scale=0.8]{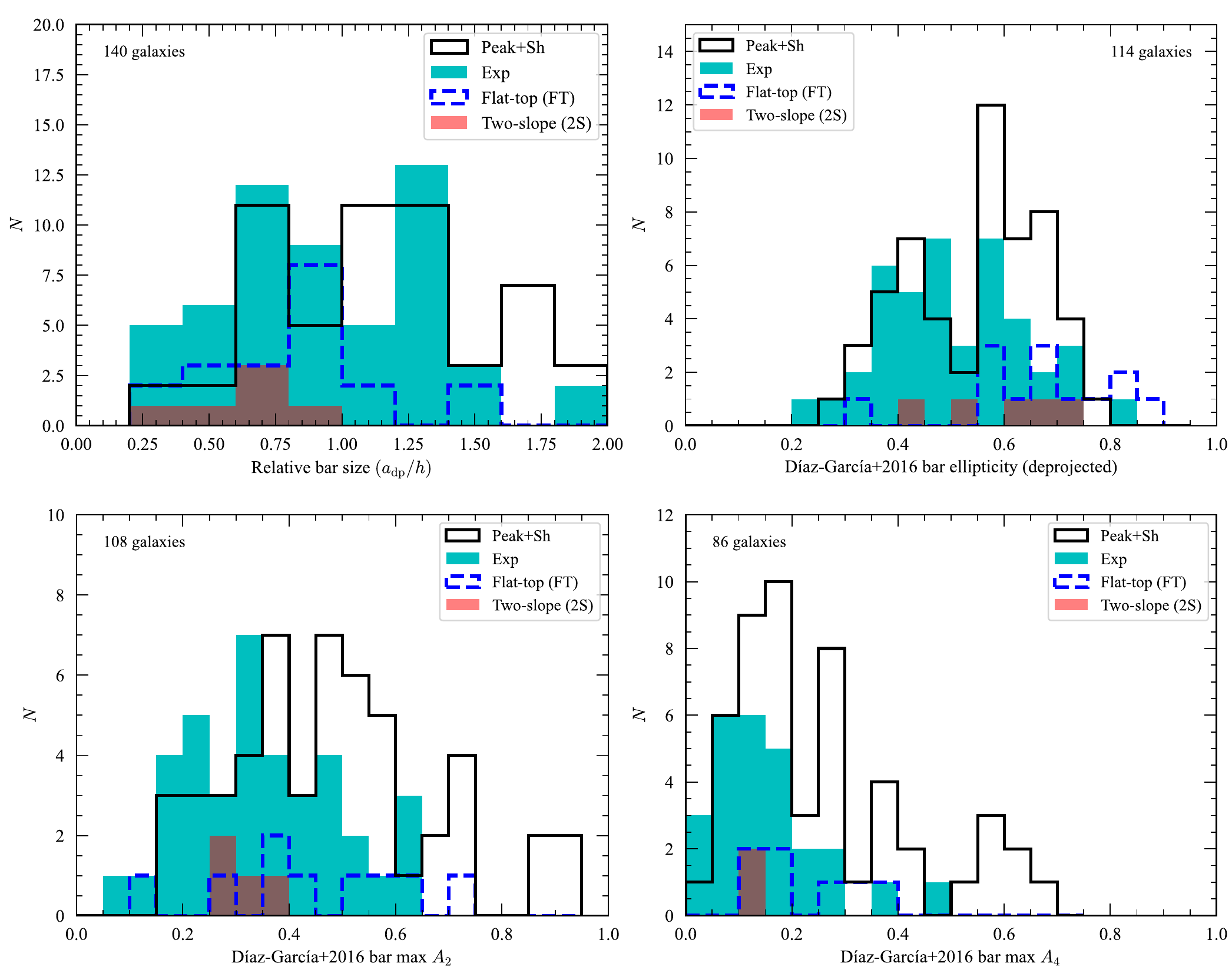}
\end{center}

\caption{As for Figure~\ref{fig:hist-profiles-vs-htype-fgas-gmr-spirals},
but now showing distribution of bar-profile classifications as a function of
relative bar size (deprojected semi-major axis divided by exponential
disc scale length, upper left) and three different measurements of bar
strength: deprojected isophotal ellipticity and maximum $m = 2$ and $m =
4$ Fourier amplitudes (relative to $m = 0$ amplitudes). The total number
of galaxies with valid measurements is indicated in the upper corners of
each plot.
\label{fig:hist-profiles-vs-barsize-barstrength-spirals} }

\end{figure*}

Finally, in
Figure~\ref{fig:hist-profiles-vs-barsize-barstrength-spirals} we show
distributions of \textit{bar} characteristics for the different profile
types. The upper left panel shows relative bar size (bar radius divided
by the exponential-disc scale length from the 2D fits of
\citealt{salo15}), while the other panels show three different
measurements of bar ``strength'', all taken from \citet{dg16a}:
deprojected maximum isophotal ellipticity of the bar, maximum $m = 2$
Fourier amplitude relative to the $m = 0$ amplitude \atwomax, and 
maximum relative $m = 4$ amplitude \afourmax.

For bar sizes, we see only a weak difference between \flatp{} and
non-\flatp{} profiles, though it is formally statistically significant
(K-S $P \sim 0.0006$, versus $P \sim 3 \times 10^{-14}$ for \logfgas{}
and $P \sim 1 \times 10^{-18}$ for Hubble type). The fact that
\flatp{} bars tend to be relatively larger is plausibly explained by
the strong stellar-mass dependence, since relative bar sizes tend to be
larger for galaxies with masses $\logmstar > 10.2$ \citep{erwin19}. 

When we turn to bar ellipticity, there is a slight tendency for \flatp{}
profiles to appear in stronger bars. However, there is considerable
overlap, so that (for example) bars with a deprojected
ellipticity of 0.4 are equally likely to have \flatp{} profiles or
Exponential profiles. In fact, when the different non-\flatp{} profiles
are combined, their deprojected ellipticity distributions basically
reproduces that of the \flatp{} galaxies (K-S test $P = 0.95$). There
is, in contrast, some evidence for a difference between the Exponential
profiles and the combination of 2S and FT profiles, in that the latter
tend to be found in more elliptical bars (K-S test $P = 0.0005$).

For \atwomax{} and \afourmax, \flatp{} profiles appear somewhat biased
toward stronger bars (higher values of \atwomax{} and \afourmax{}; K-S
$P = 0.0011$ and 0.035, respectively). The lower significance for the
\afourmax{} comparison might simply reflect the lower number of galaxies
with \afourmax{} measurements. Again, however, there is considerable
overlap, and the segregation of profile types is nowhere near as strong
as it is for stellar mass (or, indeed, for Hubble type, gas fraction,
or colour).

In summary, we find that bar-profile type is very strongly dependent on
galaxy stellar mass, in the sense that \flatp{} profiles are
overwhelmingly found in massive galaxies, while the various exponential
sub-types (Exp, FT, 2S) are found in low-mass galaxies. Similar trends
are found for Hubble type, gas mass fraction, and \gmr{} colour, with
\flatp{} profiles preferentially found in galaxies with earlier Hubble
types, lower gas fractions, and redder colours. On the other hand there
is only weak or ambiguous evidence that bar-profile class depends on
relative bar size or bar strength.

In the next section, we focus on the question of whether the observed
strong trends with stellar mass, Hubble type, gas fraction, and colour
are independent, or whether profile type might depend mostly or entirely
on just one of these characteristics, with the other trends being side
effects of known correlations between all four parameters.

\subsection{Dependence of Bar Profile Type on Single Versus Multiple Parameters}\label{sec:multiple-params}


\begin{table}
\caption{Logistic Regression for \flatp{} Profiles: Single Variables}
\label{tab:logistic}
\begin{tabular}{@{}lrrrr}
\hline
Variable & $\alpha$ & $\beta$  & \pbeta      & AIC \\
(1)      & (2)      & (3)      & (4)         & (5) \\
\hline

\hline
\multicolumn{5}{c}{Full Sample (182 galaxies)} \\

\logmstarshort   &  $-47.71$   & $4.70$   & $1.0 \times 10^{-10}$   & $112.72$ \\
Hubble type $T$   &  $2.95$   & $-0.76$   & $1.0 \times 10^{-11}$   & $140.98$ \\
\logfgas   &  $-2.84$   & $-2.48$   & $6.7 \times 10^{-10}$   & $164.67$ \\

\hline
\multicolumn{5}{c}{\bpsample{} (132 galaxies)} \\

\logvrot   &  $-34.50$   & $16.24$   & $6.7 \times 10^{-8}$   & $78.23$ \\
\logmstarshort   &  $-50.80$   & $4.99$   & $1.3 \times 10^{-7}$   & $74.79$ \\
Hubble type $T$   &  $2.71$   & $-0.70$   & $1.4 \times 10^{-8}$   & $106.95$ \\
\logfgas   &  $-2.65$   & $-2.23$   & $4.8 \times 10^{-7}$   & $122.65$ \\

\hline
\multicolumn{5}{c}{Full Sample: galaxies with \gmr{} and \atwomax{} (125 galaxies)} \\

\logmstarshort   &  $-44.25$   & $4.37$   & $2.3 \times 10^{-8}$   & $91.23$ \\
\gmr   &  $-13.19$   & $22.40$   & $5.0 \times 10^{-9}$   & $88.13$ \\
\atwomax   &  $-2.17$   & $4.02$   & $0.00021$   & $154.00$ \\
Hubble type $T$   &  $3.61$   & $-0.87$   & $2.1 \times 10^{-8}$   & $102.97$ \\
\logfgas   &  $-2.70$   & $-2.68$   & $2.8 \times 10^{-7}$   & $123.27$ \\

\hline
\multicolumn{5}{c}{\bpsample{}: galaxies with \gmr{} (118 galaxies)} \\

\logvrot   &  $-32.58$   & $15.33$   & $2.4 \times 10^{-7}$   & $70.42$ \\
\logmstarshort   &  $-48.38$   & $4.75$   & $4.7 \times 10^{-7}$   & $67.69$ \\
\gmr   &  $-12.13$   & $19.67$   & $1.8 \times 10^{-8}$   & $75.43$ \\
Hubble type $T$   &  $2.96$   & $-0.76$   & $6.4 \times 10^{-8}$   & $91.07$ \\
\logfgas   &  $-2.81$   & $-2.50$   & $1.2 \times 10^{-6}$   & $105.80$ \\

\hline
\end{tabular}

\medskip

Results of single-variable logistic regressions: probability of a barred
spiral having a \flatprof{} profile as function of values of different
parameters. Each line represents a separate logistic regression. (1)
Galaxy parameter used in fit ($\Mstar{} =$ stellar mass; $\fgas{} =$ gas
mass ratio; $\vrot =$ inclination-corrected gas rotation velocity). (2)
Intercept value for fit. (3) Slope for fit. (4) $P$-value for slope. (5)
Akaike Information Criterion value for fit; lower values indicate better
fits for a given sample.

\end{table}


\begin{table}
\caption{Logistic Regression for \flatp{} Profiles: Multiple Variables}
\label{tab:multilogistic}
\begin{tabular}{@{}lrrrr}
\hline
Variable & $\alpha$ & $\beta$  & \pbeta      & AIC \\
(1)      & (2)      & (3)      & (4)         & (5) \\
\hline

\hline
\multicolumn{5}{c}{Full Sample (182 galaxies)} \\

\logmstarshort   &  $-36.62$   & $3.80$   & $1.3 \times 10^{-6}$   & $102.99$ \\
Hubble type $T$   &   & $-0.48$   & $0.01$   &  \\
\logfgas   &   & $0.02$   & $0.98$   &  \\

\hline
\multicolumn{5}{c}{\bpsample{} (132 galaxies)} \\

\logmstarshort   &  $-42.06$   & $4.32$   & $1.4 \times 10^{-5}$   & $73.90$ \\
Hubble type $T$   &   & $-0.39$   & $0.074$   &  \\
\logfgas   &   & $0.27$   & $0.73$   &  \\

\multicolumn{5}{c}{} \\

\logvrot   &  $-24.97$   & $12.61$   & $4.3 \times 10^{-5}$   & $75.19$ \\
Hubble type $T$   &   & $-0.42$   & $0.072$   &  \\
\logfgas   &   & $-0.04$   & $0.96$   &  \\

\hline
\multicolumn{5}{c}{Full Sample: galaxies with \gmr{} and \atwomax{} (125 galaxies)} \\

\logmstarshort   &  $-33.96$   & $2.83$   & $0.0031$   & $78.99$ \\
\gmr   &   & $11.72$   & $0.035$   &  \\
\atwomax   &   & $1.51$   & $0.36$   &  \\
Hubble type $T$   &   & $-0.35$   & $0.18$   &  \\
\logfgas   &   & $0.72$   & $0.37$   &  \\

\hline
\end{tabular}

\medskip

As for Table~\ref{tab:logistic}, but now showing results of
multiple-variable logistic regressions: probability of a barred spiral
having a \flatprof{} profile as function of values of three or more
different parameters at the same time. For the \bpsample, we performed
two fits, using either \logmstar{} or \logvrot{} as a ``galaxy mass''
variable. The lines show the best-fit coefficients for the specified
variables; the first line also includes the intercept ($\alpha$) and the
AIC value for the fit. (1) Galaxy parameters used in fit. (2) Intercept
value for fit. (3) Slope for each parameter in fit. (4) $P$-value for
slope. (5) Akaike Information Criterion value for fit.

\end{table}

We have seen that \flatp{} profiles are more common in galaxies with
higher stellar masses, lower gas mass fractions, redder colours, and
earlier Hubble types. The question that naturally arises is whether
these trends are \textit{independent}, because galaxy mass is correlated
with colour and inversely correlated with gas fraction and Hubble type.
Is the dependence on these other characteristics merely a side effect of
a more general trend with stellar mass? If we could, for example,
compare barred spirals with the same stellar mass, would \flatp{}
profiles still be more common in galaxies with lower gas fractions,
redder colours, or earlier Hubble types?

We attempt to answer these questions via logistic regression, which
models the probability of a galaxy having a given binomial
characteristic as a function of one or more parameters. This is
appropriate for our problem because the visible trends show clear
segregation by parameter value: e.g., all galaxies with $\logmstar \la
9.2$ do not have \flatp{} profiles, while all galaxies with $\logmstar
\ga 10.6$ do, with the \flatp{} fraction increasing more or less
monotonically with increasing stellar mass (see
Figure~\ref{fig:fprof-BP-spirals}). Specifically, we focus on the
presence or absence of a \flatp{} profile, counting only cases where
both classifiers agreed on its presence; galaxies with split classifications
are counted as non--\flatp.

Logistic regression involves modeling the probability $P$ of a galaxy
having a particular characteristic -- e.g., a \flatp{} profile -- as
a function of one or more parameters and one or more independent
variables via the logistic equation
\begin{equation}\label{eqn-logistic}
P \; = \; \frac{1}{1 + \mathrm{e}^{-(\alpha \, + \, \sum_{i} \beta_{i} x_{i})}},
\end{equation} 
where the $x_{i}$ are the different variables (e.g., stellar mass, gas
fraction, Hubble type); the probability asymptotes to 0 as $x_{i} \rightarrow -\infty$
and to 1 as 
$x_{i} \rightarrow +\infty$ (for $\beta_{i} > 0$, with the reverse behavior
for $\beta_{i} < 0$). If a given parameter $x_{i}$ has no relation to
the \flatp{} probability, then we would expect the corresponding slope $\beta_{i}$
to be $\approx 0$. 

The best-fitting values of the parameters can be determined using a maximum
likelihood approach,\footnote{We use the standard \texttt{glm} function
in the R statistical language.} with the total likelihood being the
product of the individual Bernoulli likelihoods for each of the $N$
observed galaxies
\begin{equation}
\mathcal{L} \; = \;  \prod_{n = 1}^{N} P_{n}^{y_{n}} \, (1 - P_{n})^{1 - y_{n}} ,
\end{equation}
where $P_{n}$ is the probability for galaxy $n$ (evaluated using Eqn.~\ref{eqn-logistic}) and
$y_{n}$ is the observed result ($= 1$ if the galaxy has that characteristic
and 0 if it does not).

We can investigate the relative importance of different parameters by
performing individual fits using just one parameter and then comparing
the relative goodness of each fit. We can also fit using multiple
parameters simultaneously and look at the relative significance of
the different parameters in such fits.

\subsubsection{Dependence on Single Parameters}\label{sec:single-params-logistic}

Table~\ref{tab:logistic} compares single-variable logistic fits for
individual galaxy parameters. For each of several
subsamples,\footnote{Note that unlike the subsample counts in
Figures~\ref{fig:hist-profiles-vs-htype-fgas-gmr-spirals} and
\ref{fig:hist-profiles-vs-barsize-barstrength-spirals}, which used only
galaxies where both classifiers agreed on the final classification,
these subsamples included galaxies where both classifiers agreed on
whether a galaxy was \flatp{} or not, but may have disagreed on the
non--\flatp{} class.} we show the best-fit intercept $\alpha$ and slope
$\beta$, along with the associated probability \pbeta{} for a slope at
least that different from zero under the null hypothesis that the
true slope is zero.

We also show the Akaike Information Criterion (AIC) value
for the fits, as computed by the R function \texttt{glm}:
\begin{equation}
\mathrm{AIC} \; = \; -2 \ln \mathcal{L} \, +\,  2k, 
\end{equation} 
where $\mathcal{L}$ is the (maximized) likelihood value
(Eqn.~\ref{eqn-logistic}) and $k$ is the number of (free) parameters.
Smaller values of AIC indicate a better fit for a given data set.
Traditionally, values of $|\Delta {\rm AIC}| < 2$ are considered
non-significant, while $|\Delta {\rm AIC}| \approx 2$--6 is weak
evidence in favor of the model with lower AIC and $|\Delta {\rm AIC}| >
6$ is considered strong evidence. We note that the latter criterion
corresponds to $P < 0.05$, and so is not ``strong evidence'' by the
usual standards of astronomy; $|\Delta {\rm AIC}| \ga 12$ would be a
rough equivalent to a 3-$\sigma$ standard of significance.

This allows us to see which individual parameters are \textit{most
important}. What is clear is that the key parameters are those
related to galaxy mass, and possibly colour. For most of the samples, it
is the galaxy stellar mass that provides the best fit, although
\logvrot{} is only marginally worse. For the subset of the full sample
with colour data, it is \gmr{} (but note that this is not true for the B/P-Detection
Subsample), though with $|\Delta {\rm AIC}|$ only $\approx 4$.

While the fits using other parameters (\logfgas, \atwomax, Hubble
type) are all formally significant (i.e., they have small values of
\pbeta), they are clearly worse than galaxy mass or \gmr as predictors
(they have much larger values of AIC), and it is possible that these are
merely side-effects of correlations between, e.g., galaxy mass and the other
parameters. The question then becomes: do any of the non-mass parameters
have any \textit{independent} effect on the presence of \flatp{} profiles?

\subsubsection{Dependence on Multiple Parameters}\label{sec:multi-params-logistic}

Table~\ref{tab:multilogistic} is similar to Table~\ref{tab:logistic},
except that it shows one or two \textit{multi-parameter} logistic fits
for each subsample. Here, we look at two things: which of the parameters
in each fit have slopes that differ from zero in a statistically
significantly sense (small values of \pbeta), and which fits have
AIC values significantly smaller than the corresponding single-parameter
fits for the same subsample (Table~\ref{tab:logistic}). For example, in
the case of the full sample of 182 galaxies, the multi-parameter fit in
Table~\ref{tab:multilogistic} has AIC $\approx 103.0$ while the best
single-parameter fit in Table~\ref{tab:logistic} (using \logmstar) has
AIC $\approx 112.7$, so including the extra parameters appears to do a
better job of predicting the presence of \flatp{} profiles. Since only
the Hubble-type parameter in this multi-parameter fit has a marginally
significant ($\pbeta \approx 0.01$) slope in the multi-parameter fit,
it appears that the gas mass fraction \logfgas{} is not meaningful. 

If we restrict ourselves to the \bpsample{}, then the possible
significance of Hubble type as a secondary parameter disappears. If we
also include the \gmr{} colour and bar strength \atwomax{} (for the
125 galaxies in the full sample that have both values -- see the
final fit in the table), then \gmr{} is a marginally significant
($\pbeta \approx 0.035$) secondary parameter, while the Hubble type and
\atwomax{} are \textit{not}.

In summary, it appears that galaxy mass (\Mstar{} or \vrot) is the only
clear determinant for the presence of absence of \flatp{} profiles,
although there is the possibility that \gmr{} colour could be a
secondary parameter.

\subsection{Exponential and Related Bar Profiles} 

One of the key findings of this paper is that the classic
``exponential'' bar profile type actually consists of several distinct
sub-types. In addition to what \textit{we} call Exponential profiles
(which have a single exponential slope from the centre to the end of the
bar), we also identify Two-Slope and Flat-Top profiles
(Section~\ref{sec:profile-classif} and
Figures~\ref{fig:exp-demo}--\ref{fig:flattop-demo}).

Two-Slope bar profiles are a rarity, accounting for only
$7.1^{+2.7}_{-2.0}$\% of the non--\flatp{} bars. Flat-Top bars are
more significant, making up one quarter ($25 \pm 4$\%) of the
non--\flatp{} bars. Figures~\ref{fig:hist-profiles-vs-mstar-spirals}
and \ref{fig:hist-profiles-vs-htype-fgas-gmr-spirals} show no evidence for
any differences in stellar mass, Hubble type, or gas fraction between
Two-Slope, Flat-Top, and Exponential profiles; as noted in
Section~\ref{sec:gen-results}, K-S tests for these different sub-types
show no evidence for statistically significant differences in their
stellar-mass distributions. Further K-S tests for possible differences
in terms of Hubble type, gas mass fraction, or bar strength yielded nothing
significant, with the possible exception of a difference between Hubble
types for 2S profiles and the other two subtypes, in the sense the 2S
Hubble types are more evenly distributed and less concentrated towards
very late Hubble types ($P \sim 0.02$). The significance of this is, however,
borderline, especially considering that we are testing multiple comparisons
(e.g., 2S vs non-2S for various different galaxy parameters).

\section{The Link Between B/P Bulges and Bar Profiles}\label{sec:bp-bulges-and-profiles} 

As noted in Section~\ref{sec:intro-bp}, recent research has suggested
that ``flat'' bar profiles are associated with bars that have B/P
bulges, with the shoulder part of the profile corresponding to the
outer, vertically thin part of the bar and the B/P bulge having a
steeper surface brightness profile that produces most of the ``peak'' of
the \flatp{} profiles.

This has two obvious implications. The first is that most if not all
bars with B/P bulges should have \flatp{} profiles. The second  is
the inverse: most if not all \flatp{} profiles should be in bars with
B/P bulges. In this section of the paper, we test these ideas by
identifying which bars do and do not have B/P bulges from a
morphological perspective. To do so, we use the \bpsample{}, where we
can maximize our ability to detect both the presence and absence of B/P
bulges inside the bars.

\subsection{Determining the Frequency of B/P Bulges as a Function of Galaxy Properties}\label{sec:bp-bulges} 

\citet{erwin-debattista17} analyzed near-IR images of a local sample of
84 barred galaxies which had orientations favorable for detecting B/P
bulges. They found a very strong, almost perfectly monotonic dependence
of B/P-bulge morphology on stellar mass: galaxies with masses $<
10^{10}$ almost never had B/P bulges, while galaxies with masses $>
10^{10.5}$ almost always had them. They found supporting evidence for
this trend in the SDSS analysis of \citet{yoshino15}. Subsequently,
\citet{li17} used optical images from the Carnegie-Irvine Galaxy Survey
\citep{ho11} to argue for a very similar trend (e.g., their Figure~3),
using both the morphological signature identified by ED13 and the
``barlens'' morphology identified by Laurikainen and collaborators
\citep[e.g.,][]{laurikainen11,athanassoula15,laurikainen17}, which is
visible in more face-on galaxies. \citet{kruk19} found the same trend
using SDSS images for $z \sim 0$ barred galaxies, and hints that the
trend might be present in higher-redshift galaxies. More recently,
\citet{marchuk22} found a very similar result for B/P bulges identified
in edge-on galaxies.

The original analysis of \citet{erwin-debattista17} had some
disadvantages. In particular, it included a galaxy angular diameter limit
($D_{25} > 2\arcmin$), which translates to a bias against physically
compact galaxies. Consequently, we decided to repeat the same
morphological analysis using the \bpsample{} of our S4G-based sample,
which is strictly distance-limited and includes 132 galaxies; since
rotation velocity measurements are available for all of the
galaxies, we also look for trends in B/P fraction as a function of \vrot. We
note that \citet{erwin-debattista17} included some S0 galaxies in 
their sample, while here we are restricted to spiral galaxies only.

\subsection{Identification of B/P Bulges in Bars} 

\citet{erwin-debattista13} showed that B/P bulges in bars could be
identified via characteristic patterns in the bar isophotes, as long as
the galaxy was moderately inclined (e.g., $i \sim 40$--75\degr) and the
bar was not too close to the galaxy minor axis \citep[see
also][]{erwin-debattista16,erwin-debattista17}. This took the form of
what they called a ``box+spurs'' morphology, where the B/P bulge itself
projected to form a thick, often boxlike structure (the
``box/oval'')\footnote{In \citet{erwin-debattista13}, this was called
the ``box'', since in strong cases its isophotes were actually
rectangular; however, in many cases the projected B/P bulge has oval
isophotes, so we adopt the more general term box/oval in this paper.}
and the outer, vertically thin part of the bar projected to form
narrower isophotes (the ``spurs''). As long as the bar was oriented more
than a few degrees away from the galaxy's major axis, these two
structures were misaligned in a characteristic way, with the spurs
rotated further away from the major axis than the box/oval. (See
Figure~\ref{fig:flatbar-BP-demo} for examples, as well as figures in the aforementioned
papers.) An important corollary of this morphology was the fact that
when a B/P bulge was \textit{not} present, the bar showed symmetric,
elliptical isophotes \textit{without} the box/oval+spurs morphology,
meaning that it was possible to identify bars \textit{lacking} B/P
bulges as well.

\subsection{Updated Frequencies of B/P Bulges}\label{sec:bp-bulges-results} 

\begin{figure}
\begin{center}
\hspace*{-4mm}\includegraphics[scale=0.59]{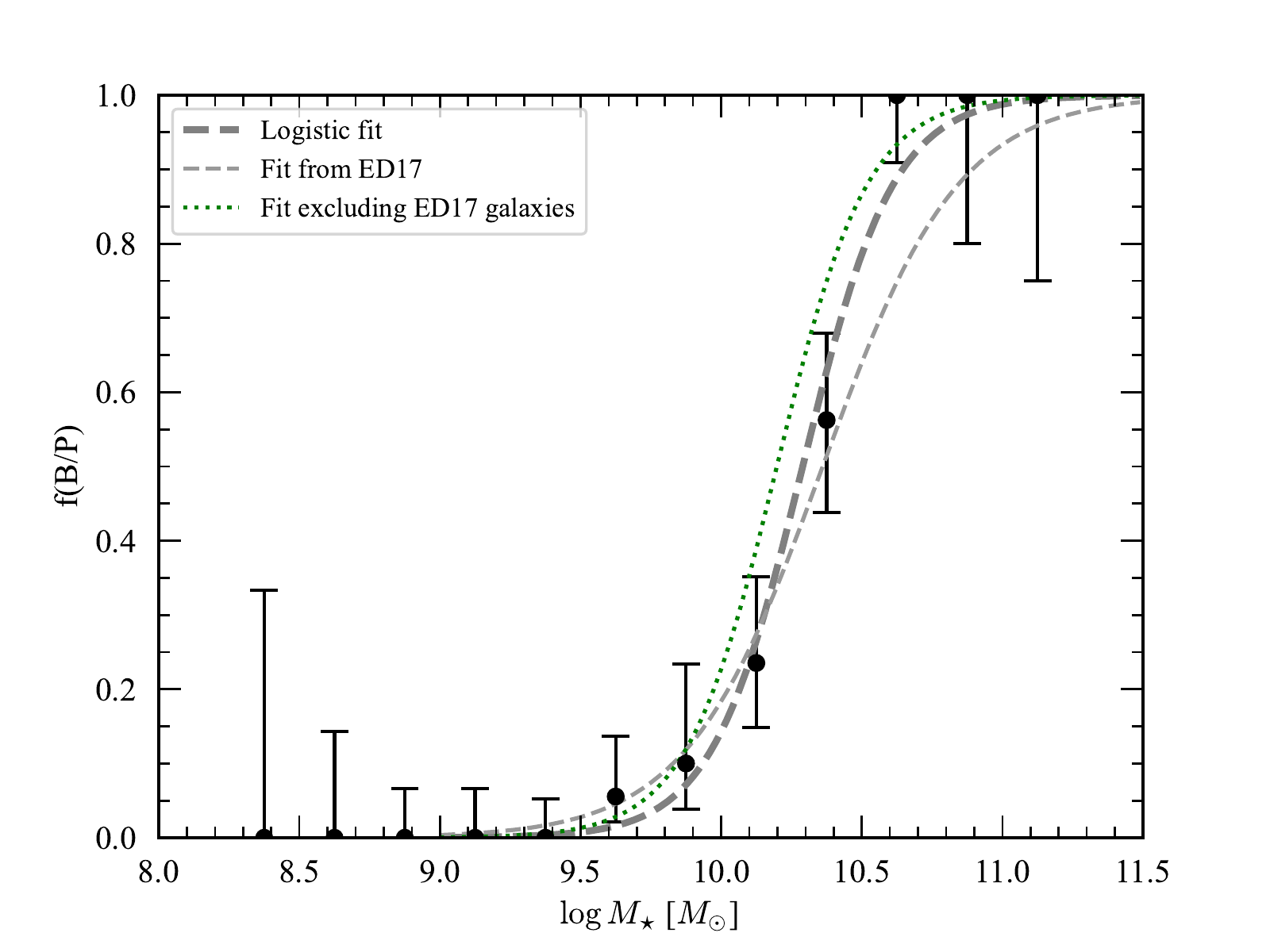}
\end{center}

\caption{Frequency of B/P bulges within bars as a function of galaxy
stellar mass, from our 132-galaxy \bpsample. The thick dashed
curve shows the best-fit logistic regression (fit to the full set of
individual data points rather than the bins). The thin dashed curve is
the logistic regression from \citet{erwin-debattista17}, while the
dotted (green) curve shows the logistic regression using the 95 galaxies in
the \bpsample{} which are \textit{not} in the original ED17 sample. 
\label{fig:fBP-mstar}}

\end{figure}

Figure~\ref{fig:fBP-mstar} shows how the fraction of bars with B/P
bulges $f(B/P)$ behaves as a function of stellar mass in the \bpsample.
The trend is a dramatic one: the B/P fraction is 0\% for $\logmstar <
9.5$ and 100\% for $\logmstar > 10.5$, with a monotonic and very steep
transition between these two regimes. The thick dashed line shows the
result of a logistic regression analysis. This indicates that $f(B/P) =
0.5$ at $\logmstar = 10.29$ and 0.9 at $\logmstar = 10.65$.

This trend is very similar to that seen for a smaller (and slightly
overlapping) sample by \citet{erwin-debattista17}, except that it is
\textit{stronger} for our newer, larger sample. (We show the original
logistic fit from \citealt{erwin-debattista17} as thin, dashed gray line,
with $f(B/P) = 0.5$ at $\logmstar = 10.36$ and 0.9 at $\logmstar =
10.89$.) We suspect this may be due to the more consistent stellar
masses used in our current sample (all based on \textit{Spitzer}
3.6\micron{} photometry); the more heterogeneous mass estimates in
\citet{erwin-debattista17} could mean a greater scatter, and thus a
weakening of the sharp transition.

There are 37 galaxies in the \bpsample{} which are also in the 84-galaxy
ED17 sample, so some of the agreement \textit{could} be due to this
overlap. We can eliminate this possibility by looking at the trend for
just those 95 galaxies in the \bpsample{} which were \textit{not} in the
ED17 sample. This shows the same basic pattern; the logistic fit for
this subsample is shown with the green line in
Figure~\ref{fig:fBP-mstar}. This increases our confidence in the general
result, since essentially the same strong B/P-fraction--stellar-mass
relation is found for two different galaxy samples.

\begin{figure}
\begin{center}
\hspace*{-2mm}\includegraphics[scale=0.57]{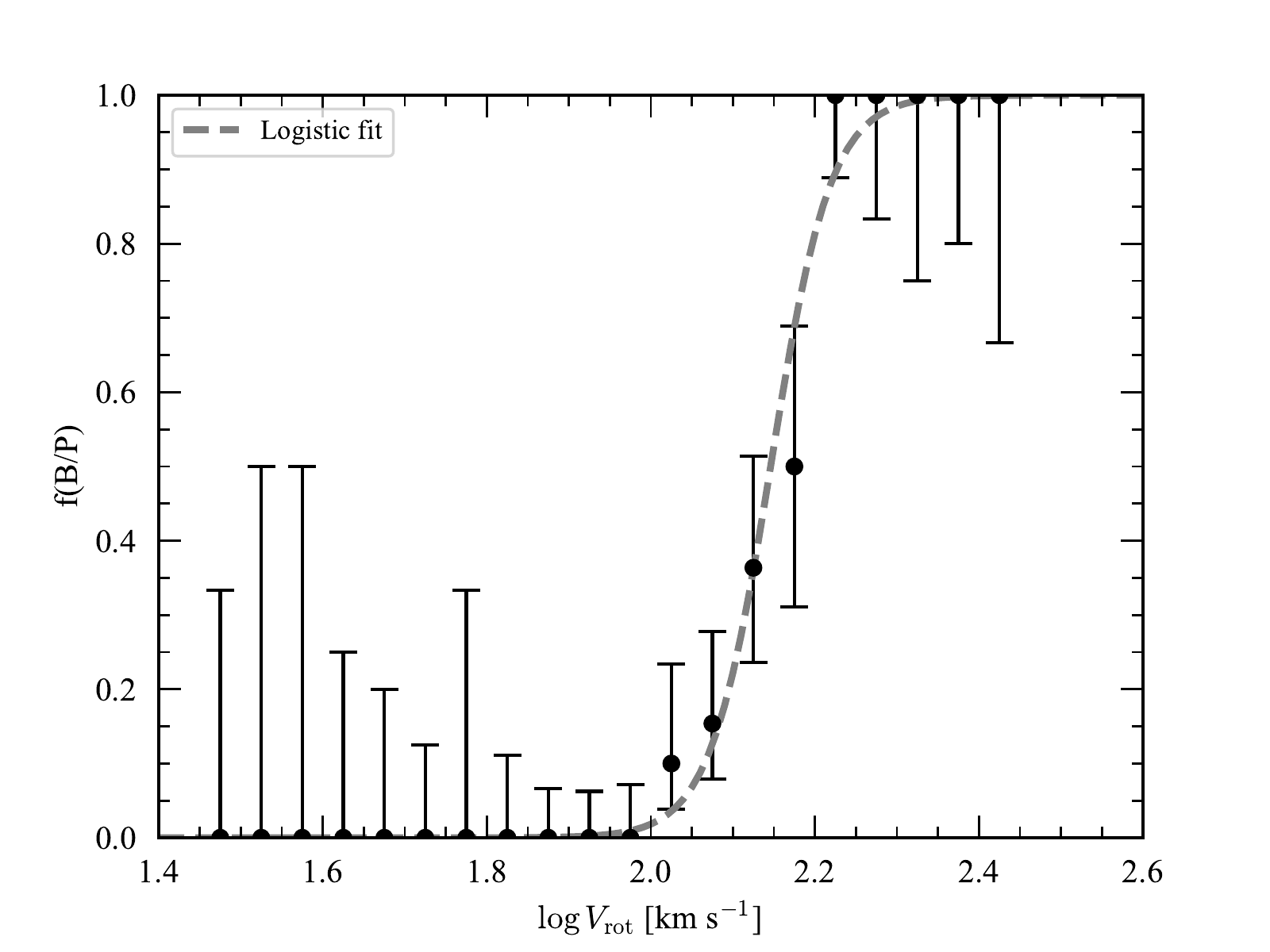}
\end{center}

\caption{As for Figure~\ref{fig:fBP-mstar}, but now showing the fraction
of bars with B/P bulges as a function of galaxy rotation velocity. 
\label{fig:fBP-vrot}}

\end{figure}

Figure~\ref{fig:fBP-vrot} shows the trend in $f(B/P)$ as a function of
galaxy (gas) rotation velocity, along with the corresponding logistic
fit. The same strong trend is clearly visible. In this case, no galaxies
with $\vrot < 100$ \kms{} have B/P bulges, while all galaxies with
$\logvrot > 2.2$ ($\vrot \sim 160$ \kms) do. The existence of the Tully-Fisher
relation means that this similarity is entirely to be expected, though
it does raise the question of whether $f(B/P)$ might somehow depend on,
e.g., halo mass rather than stellar mass. The Akaike Information
Criterion values for the logistic fits are 61.7 for the \logmstar{} fit
and 49.4 for the \vrot{} fit, which indicates the latter is a better
fit; the slope for the \vrot{} fit is clearly steeper (logistic slope $=
27.0 \pm 6.0$ versus $6.1 \pm 1.3$ for the \logmstar{} fit), indicating
a stronger trend. Of course, residual uncertainties in the stellar-mass
estimation might still introduce scatter into the latter relation, so we
cannot really conclude that halo mass is the primary driver.


\begin{table}
\caption{Logistic Regression for Presence of B/P Bulges}
\label{tab:bp-logistic}
\begin{tabular}{@{}lrrrr}
\hline
Variable        & $\alpha$ & $\beta$  & \pbeta & AIC \\
(1)             & (2)      & (3)      & (4)         & (5) \\
\hline

\logmstarshort   &  $-63.62$   & $6.18$   & $3.7 \times 10^{-6}$   & $61.85$ \\
\logvrot   &  $-58.51$   & $27.26$   & $1.2 \times 10^{-5}$   & $48.12$ \\

\hline
\end{tabular}

\medskip

Results of single-variable logistic regressions: probability of a barred
spiral having a B/P bulge as function of stellar mass or rotation velocity
for the \bpsample. 
(1) Galaxy parameter used in fit (\Mstar{} = stellar mass; \vrot{} = gas
rotation velocity). (2) Intercept value for fit. (3) Slope for fit. (4)
$P$-value for slope. (5) Akaike Information Criterion value for fit.

\end{table}

\subsection{\flatprof{} Profiles and the Presence of B/P Bulges}
\label{sec:flatprof-bp} 

We now examine possible connections between B/P bulges and bar profiles.
We first look at bars in the \bpsample{} with identified B/P bulges (as
a reminder, this sample is made of galaxies with inclinations and bar
orientations selected to maximize the ability to detect B/P bulges, if
they are present). Essentially \textit{all} such bars have \flatp{}
profiles: we securely classify 30 of the 32 B/P-bulge hosts as
having \flatp{} profiles -- and for the two galaxies not unambiguously
classified (NGC~4498 and NGC~7513), the classifications were split, with
one of the two classifiers labeling them \flatp.

What about the inverse? Do we find \flatp{} profiles in galaxies
whose bars \textit{lack} B/P bulges? There are in fact nine galaxies in
the \bpsample{} which both classifiers deemed to have \flatp{}
profiles, but for which we found little or no evidence for B/P bulges.
(The two classifiers differed on four other galaxies without B/P bulges,
with one \flatp{} and one non--\flatp{} classification for each.)
Thus, it appears that \flatp{} profiles \textit{can} sometimes occur
in the absence of B/P bulges.

\begin{figure}
\begin{center}
\hspace*{-3.5mm}\includegraphics[scale=0.58]{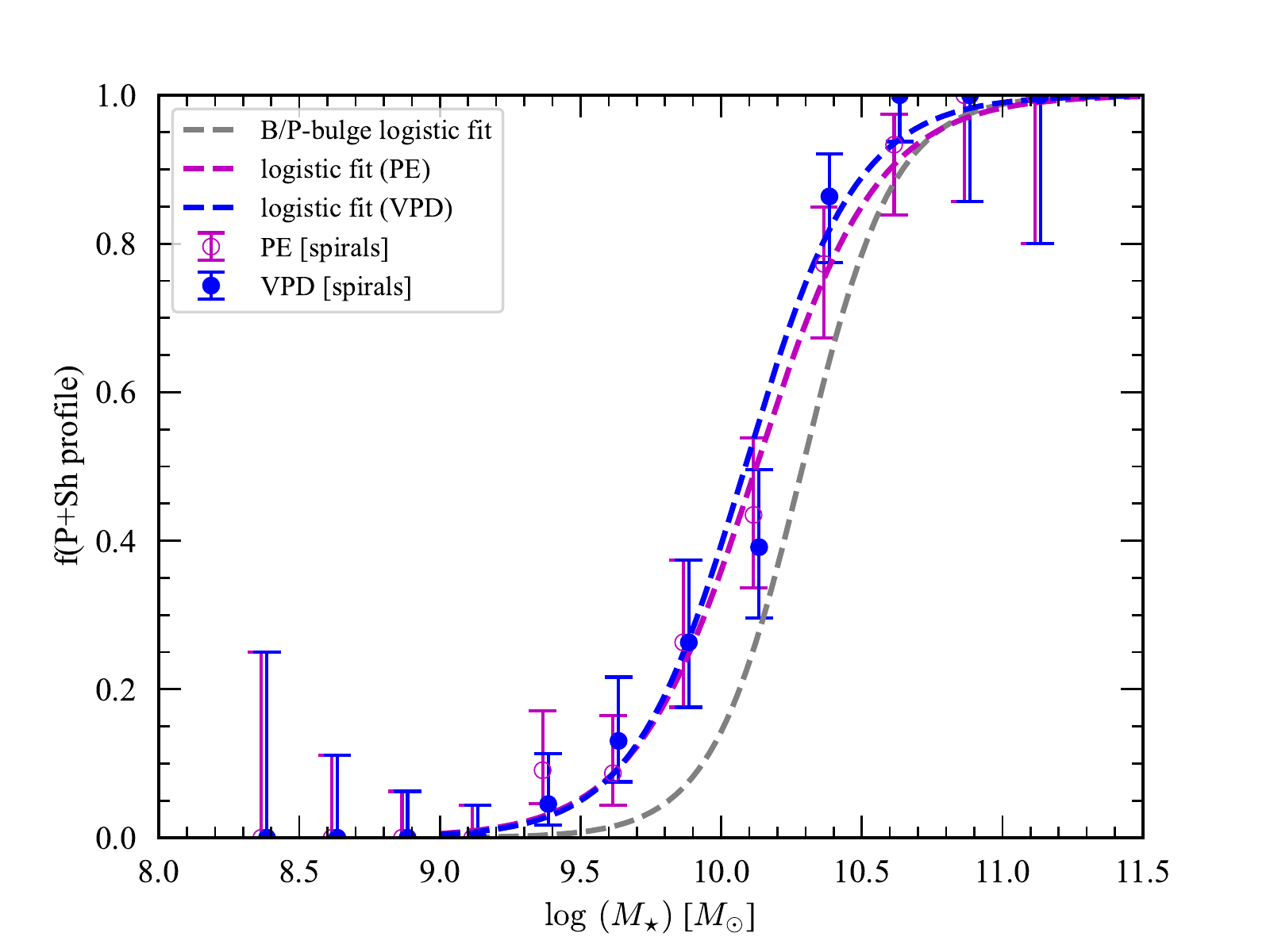}
\end{center}

\caption{Frequency of \flatp{} bar profiles in barred spiral galaxies
(both subsamples combined) as a function of stellar mass; different symbols indicate 
results from the two classifiers. Dashed coloured curves show corresponding logistic fits.
The grey dashed curve is the logistic fit for B/P-bulge 
presence from Figure~\ref{fig:fBP-mstar}.\label{fig:fprof-BP-spirals}}

\end{figure}

The strong mass dependence seen for B/P bulges
(Subsection~\ref{sec:bp-bulges-results}) is replicated in the strong
mass dependence of \flatp{} profiles, as can be seen in
Figure~\ref{fig:fprof-BP-spirals}. There is a clear implication: either
one of these things \textit{causes} the other (e.g., the buckling
instability that gives rise to B/P bulges also produces \flatp{} bar
profiles), or both are linked to some common underlying mechanism. The
fraction of bars with \flatp{} profiles is slightly \textit{higher} than
the B/P-bulge fraction at all masses (that is, at all masses where the
fractions are $> 0$ and $< 1$). The logistic fits suggest that the
\flatp{} fraction reaches 50\% at $\logmstar \approx 10.1$, while the
B/P-bulge fraction reaches 50\% at $\logmstar \approx 10.3$ (compare the
plotted curves in Figure~\ref{fig:fprof-BP-spirals}). A K-S test gives
$P = 0.0059$ for the null hypothesis that B/P hosts with \flatp{}
profiles and \textit{non}-B/P hosts with \flatp{} profiles come from the
same parent distribution of stellar masses, so there is some suggestion
of a real difference in the stellar-mass distributions.\footnote{The
median stellar mass of the non-B/P \flatp{} galaxies is $\logmstar \sim
10.2$, while for the \flatp{} galaxies with B/P bulges it is $\sim
10.6$.} If bars form earlier and/or evolve faster in higher-mass
galaxies, then this might be an indication that formation of the
\flatp{} profiles \textit{precedes} formation of B/P bulges (see
Section~\ref{sec:timing}), which could perhaps explain why the mass
distribution of B/P hosts and non-B/P hosts with \flatp{} profiles
differs slightly.

\section{Discussion}\label{sec:discuss} 

\subsection{Comparison with Previous Studies}

Our results are generally quite consistent with the original work by
\citet{ee85}, \citet{elmegreen96}, and \citet{regan97}: we find a strong
segregation of profile type with Hubble type, in the sense that
\flatp{} profiles (corresponding to the earlier ``flat'' profile
type) are preferentially found in earlier-type spirals and non-\flatp{}
profiles (corresponding to the earlier ``exponential'' type) are
preferentially found in late-type spirals. However, we argue on the
basis of our multi-parameter logistic fits
(Section~\ref{sec:multiple-params}) that the Hubble-type dependence is
basically a side effect of a more fundamental dependence on galaxy
mass.

The studies of \citet{kim15} and \citet{kruk18} both suggested a strong
dependence of bar-profile type -- parametrized via the S\'erisc index of
the bar component in their 2D image decompositions  -- on stellar mass.
\citet{kim15} argued for a  transition between ``flat'' (bar-component
S\'ersic $n < 0.4$) and ``exponential'' ($n \geq 0.8$) occurring at
$\logmstar \approx 10.2$. This is very similar to our transition mass of
$\logmstar \approx 10.1$, defined as the point where the fraction of
bars with \flatp{} profiles reaches 50\% in our logistic regression
(Section~\ref{sec:flatprof-bp}). \citet{kruk18} found a difference in
S\'ersic bar index between their low-mass bin ($\logmstar < 10.25$),
where $n_{\rm bar} = 0.81 \pm 0.60$, and their high-mass bin ($\logmstar
\ge 10.25$), where $n_{\rm bar} = 0.43 \pm 0.47$.

Both \citet{kim15} and \citet{kruk18} also pointed to associations
between $B/T$ values and bar-profile types. For example, Kim et al.\
argued, ``The majority of exponential bars are in bulgeless galaxies,
and all galaxies with $n_{\rm bar} > 0.7$ are bulgeless galaxies. Thus,
bar profiles can be better separated by bulge dominance and bulge types
than by galaxy mass.'' Similarly, Kruk et al.\ noted that low values of
the bar S\'ersic index (corresponding to flatter [outer] bar profiles)
were strongly associated with ``obvious bulge'' galaxies (i.e., those
that required an extra S\'ersic component for the ``bulge'' in their
image decompositions), while more exponential-like bar profiles were
associated with ``disc dominated'' galaxies.

Although we do not consider $B/T$ values for our galaxies, the
bulge--bar-profile associations of \citet{kim15} and \citet{kruk18} can
be understood in the context of our findings if we recall that our
\flatp{} profile is the \textit{combination} of a shallow-to-flat outer
profile and a steep inner profile. Thus, a \flatp{} bar is naturally
best represented in a 2D fit by the combination of a low-$n$ S\'ersic
component for the outer part of the bar and an additional, smaller, and
higher-$n$ S\'ersic component for the inner part of the profile. Our
argument is that the steep, inner part of the bar's profile should
\textit{not} be seen as a separate (spheroidal) ``bulge'', but rather as
the combination of the bar's B/P structure and any additional inner,
non-bar components (nuclear discs, nuclear bars, actual spheroids, etc.)
that may be present.\footnote{We remind the reader that there is some
evidence for a small subset of bars with \flatp{} profiles but no B/P
bulge (Section~\ref{sec:flatprof-bp}).} Similar arguments have been made
in the context of 2D image fitting of bars by \citealt{neumann19} and
\citealt{erwin21}.

\citet{lee19} suggested that flat bar profiles are significantly more
common in SB bars than in SAB bars. In our sample, we find no evidence
for this -- in fact, the opposite is true, though not at a statistically
significant level. The fact that Lee et al.\ found over one-third of
their \textit{unbarred} galaxies to have flat profiles suggests their
method of deriving profiles (combining amplitudes from Fourier analyses
of deprojected images) may not produce the same results as ours.
 
Finally, we have argued that ``exponential'' bar profiles really fall into three
subclasses: single-exponential (Exp), two-slope (2S), and flat-top (FT).
This raises a question: why were 2S and FT profiles not identified in
earlier studies? A perusal of \citet{ee85}, \citet{elmegreen96}, and
\citet{regan97} shows only ten definite exponential profiles. Since the
profiles from \citet{ee85} were from centrally saturated images, making
it difficult or impossible to identify central deviations from a
single-exponential profile, we concentrate on the seven
exponential-profile galaxies from \citet{elmegreen96} and
\citet{regan97}.\footnote{\textit{Spitzer} IRAC1 images indicate that NGC~3184 is
not actually barred \citep[e.g.,][]{buta15,herrera-endoqui15}, so we
exclude this galaxy.}  Two of these were classified as part of our
sample; we performed similar classifications for the other five galaxies
(NGC~925, NGC~1359, NGC~1744, NGC~2835, and NGC~7741).

Of these seven ``classical exponential'' profiles from the literature,
we find one to be ambiguous (\flatp{} according to one classifier,
Exponential according to the other), with the remaining six are
Exponential profiles. Since 68\% of the non-\flatp{} bars in our sample
are Exponential rather than 2S or FT, it is plausible that previous
studies have missed the presence of FT and 2S bar profiles due to
small-number statistics.

\subsection{Comparison with Theory}

As noted in the Introduction (Section~\ref{sec:intro-trad}), theoretical work
attempting to explain the origin of bar profiles has been relatively
scant. In this section, we address some of what has been done in this area.

\citet{athanassoula02a} noted differences in bar profiles in three
different $N$-body simulations. The clearest case of a \flatp{} profile --
indeed, probably the first instance of a simulation producing a proper
\flatp{} profile -- 
was in their MH (``massive halo'') model, while the MD (``massive
disc'') model had a bar with an approximately exponential
profile. Unfortunately, since we have no kinematic information for the vast
majority of our galaxies, we cannot test whether the observed profiles
might correlate with different halo-density profiles or different relative
halo-disc concentrations.

\citet{anderson22} analyzed an ensemble of barred-galaxy simulations
(mostly $N$-body, but including three with gas and star formation) using
an algorithm for detecting and measuring the presence of shoulders in
the surface-density profiles of bar major axes. They also checked for
bar buckling and the presence or absence of B/P bulges (this included
cases where B/P bulges formed gradually \textit{without} strong
buckling).

They identified numerous instances of shoulders -- what we would term
\flatp{} profiles -- and found that the shoulders were primarily made up
of particles trapped around looped $x_{1}$ orbits. The formation of
shoulders appeared to be tied to the secular evolution of the bar,
and in particular of the \textit{growth} of the bar, so that shoulders
did not appear when the bar did not grow.

Since the focus of \citet{anderson22} was on \flatp{} profiles, the different
instances of \textit{not}-\flatp{} profiles were not classified. Their
Figs.~7 and 8 show that such profiles tended to be pure exponential.
However, in a few cases the profile showed a Flat-Top shape, usually
very early on (i.e., immediately after bar formation), and then evolved
into an Exponential profile (and often then on to \flatp{}).

They noted that B/P bulges and \flatp{} profiles were commonly but
not always associated in their simulations, including cases where B/P
bulges formed \textit{before} shoulders \textit{and} cases where secondary
buckling eliminated shoulders, transforming (or returning) the profile
to exponential. This is in partial disagreement with our observational
findings: while we do find some bars with \flatp{} profiles
that do not have B/P bulges, we find no instances of the opposite.
One possible implication might be that secondary buckling, which
should result in B/P bulges without \flatp{} profiles, is quite rare
in real galaxies.

\subsection{Timing Considerations}\label{sec:timing}

In Section~\ref{sec:flatprof-bp}, we noted that while all bars with
B/P bulges show \flatp{} profiles, there are some bars with \flatp{}
profiles that do not seem to have B/P bulges. This suggests a possible
scenario where \flatp{} profiles form first, before the appearance of
B/P bulges. In this section, we make some crude duty-cycle estimates to
see if we can relate the observed frequencies to possible timescales --
in particular, we sketch out an estimate of what fraction of the time
bars might spend with \flatp{} profiles before forming B/P bulges.

For simplicity, we assume that bars form (or re-form) at some
uniform rate in time $R$, and that they have a finite lifetime $L$; the
number of barred galaxies is then $N_{\rm bar} = R L$. (If bars have
lifetimes longer than the current age of the universe, then $L$ is the
time since bars started forming.) We also assume that there is a delay
time $T_{1}$ between bar formation and the formation of the \flatp{}
profile, and a further delay time $T_{2}$ between \flatp{} profile
formation and the formation of an observable B/P bulge. Thus, the number
of barred galaxies with \flatp{} profiles is $N_{\rm \flatp} = R (L -
T_{1})$ and the number of barred galaxies with both \flatp{} profiles
and B/P bulges is $N_{\rm both} = R (L - T_{1} - T_{2})$, so the number
of barred galaxies with \flatp{} profiles but \textit{without} B/P
bulges is just $N_{\rm \flatp-only} = R T_{2}$. This lets us see that
the ratio of galaxies with \flatp{} but no B/P bulges to all barred
galaxies is $N_{\rm \flatp-only} / N_{\rm bar} = T_{2}/L$.

For our sample, $N_{\rm \flatp-only} / N_{\rm bar} \approx 0.068$, 
which implies that the typical time between \flatp{} formation and B/P 
formation is $\sim 0.07$ of the bar lifetime. If bars are permanent (or 
at least have lifetimes longer than the current age of the universe) and 
began forming $\sim 10$ Gyr ago \citep[e.g.,][]{guo23}, then the time 
between \flatp{} formation and B/P formation would be $\sim 0.7$ Gyr.

\section{Summary}\label{sec:summary} 

We have presented an analysis of volume- and mass-limited samples of
barred spiral galaxies (excluding lenticulars) wherein we revisit the
classic question, first raised by \citet{ee85}, of what forms the
major-axis surface-brightness profiles of bars can take. We argue that
the classic dichotomy first reported by \citet{ee85} -- that bar
profiles fall into ``flat'' and ``exponential'' types -- is better
understood in terms of \textit{four} profile types: 
\begin{enumerate}

\item \flatprof{} (\flatp): This is an update of the traditional flat
bar profile. In such bars, the inner part of the profile forms a steep
central \textit{peak} (in combination with additional, non-bar structures
such as classical bulges, nuclear discs, secondary bars, etc.), while
the outer part of the profile forms a \textit{shoulder}: a shallower
(sometimes actually flat) subsection with an outer break to a steeper
falloff.

\item Exponential: This a bar-major-axis profile that is
essentially a pure exponential extending into the centre of the galaxy
(ignoring local variations due to star formation and nuclear
star clusters).

\item Two-Slope (2S): This is a profile with a shallow inner exponential
and a steeper outer exponential.

\item Flat-Top (FT): Here, the inner part of the bar profile is approximately
flat (i.e., constant surface brightness), with a steeper falloff in the
outer part of the bar. 
\end{enumerate}

The three non-\flatp{} profile types (Exponential, 2S, and FT, with
Exponential being the most common) are effectively subsets of the
original exponential type of \citet{ee85}. Our subdivision of this type
comes from the fact that we consider the entire bar-major-axis profile,
extending into the centre of the galaxy, rather than just the outer part
of the bar as in most previous studies; this enables us to see
differences in the inner parts of bars that have otherwise similar outer
profiles.

We find a very strong mass segregation for the different profile types,
in the sense that \flatp{} bars are found in high-mass galaxies, while
the other types are found in low-mass galaxies (with no clear difference
between their distributions). This is consistent with the original study
of \citet{ee85}, which found flat profiles in early-type spirals and
exponential profiles in late-type spirals, when one takes into
account the fact that late-type spirals are generally lower in mass than
early-type spirals, and thus more prone to host bars in the
``exponential sub-family'' (Exponential, Two-slope, or Flat-Top).

\flatp{} and non-\flatp{} bars are also segregated by (neutral) gas
fraction, global galaxy colour, and (weakly) bar strength, with \flatp{}
bars preferentially found in gas-poor and redder galaxies, and those
with stronger bars. A careful analysis shows, however, that these trends
(as well as the Hubble-type trend) are mostly if not entirely
\textit{side effects} of the dominant mass-segregation trend: there is
no clear evidence for any dependence of bar-profile type on Hubble
type or gas fraction, once the dependence on mass is controlled for, and
only weak evidence for a possible additional dependence on colour.
Additionally, we find no evidence for a systematic difference in
(relative) bar size or strength between \flatp{} and non-\flatp{} bars
once stellar mass is accounted for.

As part of our analysis, we classify a subsample of bars in an
inclination- and bar-position-angle-limited subsample (\bpsample) into
those with and without B/P bulges inside their bars. In line with
previous work \citep{erwin-debattista17,li17,marchuk22}, we find a very strong
dependence of B/P presence on galaxy stellar mass (or, equivalently,
gas rotation velocity \vrot); this dependence is even stronger than
found for the smaller sample of \citet{erwin-debattista17}, possibly
because we use a more consistent set of stellar mass estimates in 
this paper.

We find a near-perfect match between bars with B/P bulges and the
\flatp{} class: when a bar has an identifiable B/P bulge, its major-axis
profile is \flatp{}, with the peak being due to the steep profile of the
B/P bulge (plus any extra, non-bar components near the centre) and the
shoulders associated with the vertically thin outer part of the bar.
There is a small population of bars lacking B/P bulges which
nevertheless have \flatp{} profiles, mostly at intermediate stellar
masses ($\logmstar \sim 10.2$); this may be a hint that formation of
\flatp{} profiles \textit{precedes} the formation of B/P bulges.

\section*{Acknowledgments} 

We would like to thank the referee, Bruce Elmegreen, for useful comments
and questions, as well as Lia Athanassoula, Leandro Beraldo e Silva,
Adriana de Lorenzo-C{\'a}ceres, Martin Herrera-Endoqui, Johan Knapen, Zhao-Yu
Li, and Jerry Sellwood for various helpful comments and suggestions. We also
thank the anonymous author of  \textit{Sir Gawain and the Green Knight}
for certain bits of inspiration.

This work is based in part on observations made with the
\textit{Spitzer} Space Telescope, obtained from the NASA/IPAC Infrared
Science Archive, both of which are operated by the Jet Propulsion
Laboratory, California Institute of Technology under a contract with the
National Aeronautics and Space Administration. This paper also makes use of
data obtained from the Isaac Newton Group Archive which is maintained as
part of the CASU Astronomical Data Centre at the Institute of Astronomy,
Cambridge.

The Legacy Surveys consist of three individual and complementary
projects: the Dark Energy Camera Legacy Survey (DECaLS; Proposal ID
\#2014B-0404; PIs: David Schlegel and Arjun Dey), the Beijing-Arizona Sky
Survey (BASS; NOAO Prop. ID \#2015A-0801; PIs: Zhou Xu and Xiaohui Fan),
and the Mayall z-band Legacy Survey (MzLS; Prop. ID \#2016A-0453; PI:
Arjun Dey). DECaLS, BASS and MzLS together include data obtained,
respectively, at the Blanco telescope, Cerro Tololo Inter-American
Observatory, NSF’s NOIRLab; the Bok telescope, Steward Observatory,
University of Arizona; and the Mayall telescope, Kitt Peak National
Observatory, NOIRLab. The Legacy Surveys project is honoured to be
permitted to conduct astronomical research on Iolkam Du’ag (Kitt Peak),
a mountain with particular significance to the Tohono O’odham Nation.

NOIRLab is operated by the Association of Universities for Research in
Astronomy (AURA) under a cooperative agreement with the National Science
Foundation.

This project used data obtained with the Dark Energy Camera (DECam),
which was constructed by the Dark Energy Survey (DES) collaboration.
Funding for the DES Projects has been provided by the U.S. Department of
Energy, the U.S. National Science Foundation, the Ministry of Science
and Education of Spain, the Science and Technology Facilities Council of
the United Kingdom, the Higher Education Funding Council for England,
the National Center for Supercomputing Applications at the University of
Illinois at Urbana-Champaign, the Kavli Institute of Cosmological
Physics at the University of Chicago, Center for Cosmology and
Astro-Particle Physics at the Ohio State University, the Mitchell
Institute for Fundamental Physics and Astronomy at Texas A\&M
University, Financiadora de Estudos e Projetos, Funda{\c{c}}{\~a}o
Carlos Chagas Filho de Amparo, Financiadora de Estudos e Projetos,
Funda{\c{c}}{\~a}o Carlos Chagas Filho de Amparo {\`a} Pesquisa do
Estado do Rio de Janeiro, Conselho Nacional de Desenvolvimento
Cient{\'i}fico e Tecnol{\'o}gico and the Minist{\'e}rio da Ci{\^e}ncia,
Tecnologia e Inova{\c{c}}o{\~e}s, the Deutsche Forschungsgemeinschaft and the
Collaborating Institutions in the Dark Energy Survey. The Collaborating
Institutions are Argonne National Laboratory, the University of
California at Santa Cruz, the University of Cambridge, Centro de
Investigaciones Energ{\'e}ticas, Medioambientales y
Tecnol{\'o}gicas-Madrid, the University of Chicago, University College
London, the DES-Brazil Consortium, the University of Edinburgh, the
Eidgen{\"o}ssische Technische Hochschule (ETH) Z{\"u}rich, Fermi
National Accelerator Laboratory, the University of Illinois at
Urbana-Champaign, the Institut de Ciencies de l’Espai (IEEC/CSIC), the
Institut de F{\'i}sica d’Altes Energies, Lawrence Berkeley National
Laboratory, the Ludwig Maximilians Universit{\"a}t M{\"u}nchen and the
associated Excellence Cluster Universe, the University of Michigan,
NSF’s NOIRLab, the University of Nottingham, the Ohio State University,
the University of Pennsylvania, the University of Portsmouth, SLAC
National Accelerator Laboratory, Stanford University, the University of
Sussex, and Texas A\&M University.

BASS is a key project of the Telescope Access Program (TAP), which has
been funded by the National Astronomical Observatories of China, the
Chinese Academy of Sciences (the Strategic Priority Research Program
``The Emergence of Cosmological Structures'' Grant \# XDB09000000), and the
Special Fund for Astronomy from the Ministry of Finance. The BASS is
also supported by the External Cooperation Program of Chinese Academy of
Sciences (Grant \# 114A11KYSB20160057), and Chinese National Natural
Science Foundation (Grant \# 11433005).

The Legacy Survey team makes use of data products from the Near-Earth
Object Wide-field Infrared Survey Explorer (NEOWISE), which is a project
of the Jet Propulsion Laboratory/California Institute of Technology.
NEOWISE is funded by the National Aeronautics and Space Administration.

The Legacy Surveys imaging of the DESI footprint is supported by the
Director, Office of Science, Office of High Energy Physics of the U.S.
Department of Energy under Contract No. DE-AC02-05CH1123, by the
National Energy Research Scientific Computing Center, a DOE Office of
Science User Facility under the same contract; and by the U.S. National
Science Foundation, Division of Astronomical Sciences under Contract No.
AST-0950945 to NOAO.

The Siena Galaxy Atlas was made possible by funding support from the
U.S. Department of Energy, Office of Science, Office of High Energy
Physics under Award Number DE-SC0020086 and from the National Science
Foundation under grant AST-1616414.

This research also made use of Astropy, a community-developed core Python
package for Astronomy \citep{astropy22}.

\section*{Data availability}

The data underlying this article, along with code for reproducing fits
and figures, are available at
\url{https://doi.org/10.5281/zenodo.7545179}. (\textit{Spitzer}
images for all galaxies can be found at, e.g., the NASA
Extragalactic Database: \url{https://ned.ipac.caltech.edu}.)


\bibliographystyle{mnras}

\appendix{}

\section{Bar-Profile Plots Used for Blind Classifications}\label{app:plots-for-classif}

\begin{figure*}
\begin{center}
\hspace*{-3.5mm}\includegraphics[scale=0.95]{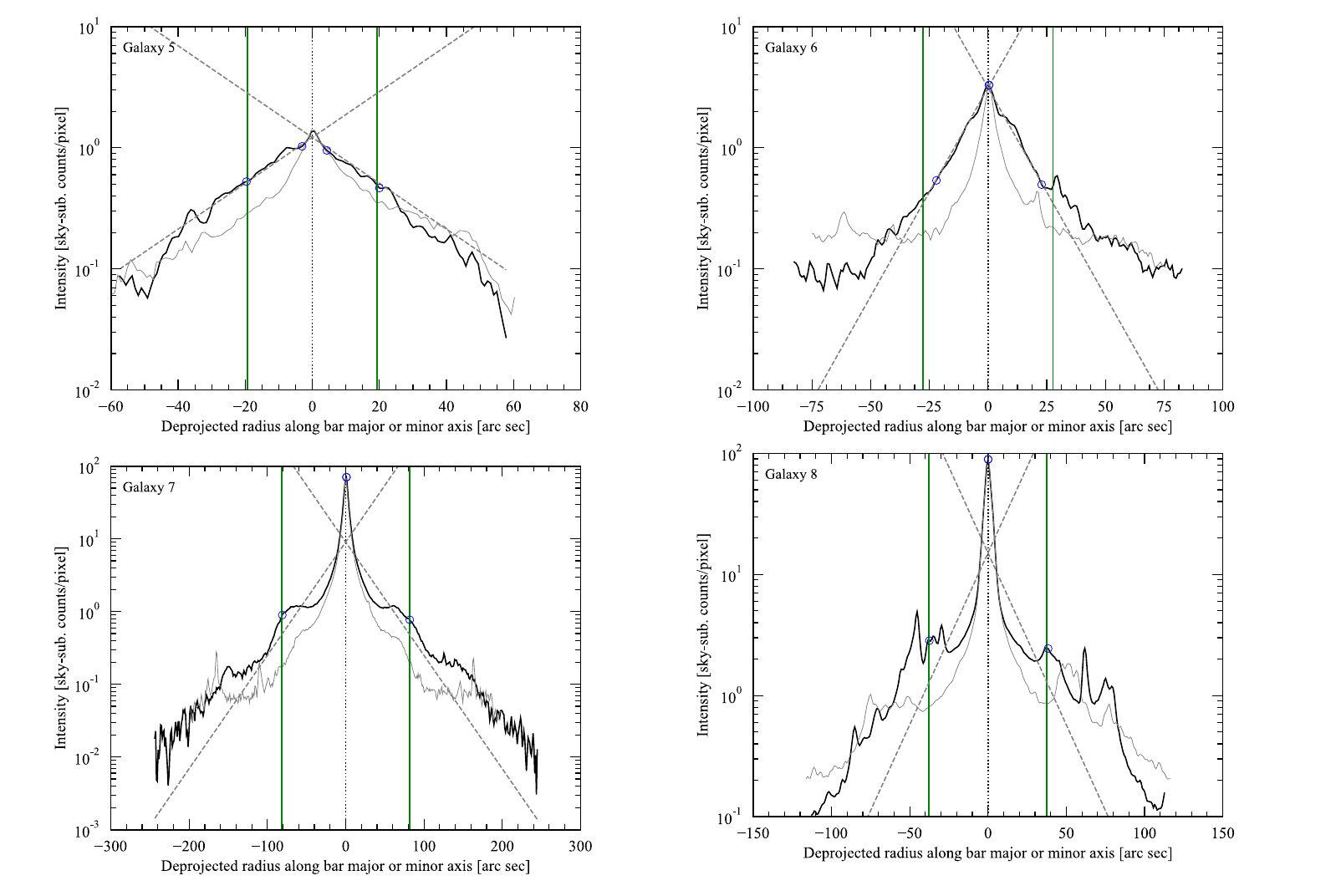}
\end{center}

\caption{Examples of the profile plot format used for our blind classifications. Solid lines
indicate the bar-major-axis profile while thin grey lines show the bar-minor-axis profile.
Vertical green lines mark the bar radius. Dashed diagonal lines show simple
exponential fits to the bar-major-axis profiles, using data between the open
circles. \label{fig:profile-plots-for-classif}}

\end{figure*}

Figure~\ref{fig:profile-plots-for-classif} shows examples of the actual
profile plots used for our classification. To evaluate the possible
exponential nature of the profiles, we plotted an automatic exponential
fit to the profile from the centre to the bar radius (dashed lines in
Figure~\ref{fig:profile-plots-for-classif}). In the top two panels, the
bar profiles match quite well with the exponential fits, and we would
classify both profiles as Exponential. The bottom two plots show strong
\flatp{} profiles, with the outer parts of the shoulders protruding
above the exponential fit, the inner shoulder and the outer part of the
peak lying clearly below the fit, and the central peak standing well
above the extrapolation of the fit to $r = 0$.

\bsp	
\label{lastpage}
\end{document}